\documentclass[fleqn,usenatbib]{mnras}

\usepackage{newtxtext,newtxmath}
\usepackage{hyperref}

\usepackage[T1]{fontenc}

\DeclareRobustCommand{\VAN}[3]{#2}
\let\VANthebibliography\thebibliography
\def\thebibliography{\DeclareRobustCommand{\VAN}[3]{##3}\VANthebibliography}

\newcommand{\CMD}{colour-magnitude diagram}

\usepackage{graphicx} 
\usepackage{amsmath} 
\usepackage{tikz}

\title[Exploring the IPMO population in USco]{
Exploring the planetary-mass population in the Upper Scorpius association
\thanks{Based on observations collected at the European Organisation for Astronomical Research 
in the Southern Hemisphere under ESO programmes 089-C.0102(ABC), 097.C-0781(A), and 0101.C-0565.}
}

\author[N. Lodieu et al.]{
N.\ Lodieu $^{1,2}$\thanks{E-mail: nlodieu@iac.es},
N.\ C.\ Hambly $^{3}$,
N.\ J.\ G.\ Cross $^{3}$
\\
$^{1}$Instituto de Astrof\'isica de Canarias (IAC), C/ V\'ia L\'actea s/n,
E-38200 La Laguna, Tenerife, Spain \\
$^{2}$Departamento de Astrof\'isica, Universidad de La Laguna (ULL),
E-38206 La Laguna, Tenerife, Spain \\
$^{3}$Scottish Universities Physics Alliance (SUPA), Institute for Astronomy, School of Physics and Astronomy, University of Edinburgh, \\ Royal Observatory, Blackford Hill, Edinburgh EH9 3HJ, UK
}

\date{Accepted \today. Received \today; in original form \today}

\pubyear{2015}

\begin{document}
\label{firstpage}
\pagerange{\pageref{firstpage}--\pageref{lastpage}}
\maketitle

%
%
\begin{abstract}
We aim at identifying very low-mass isolated planetary-mass member candidates in the nearest OB association to the Sun, Upper Scorpius (145 pc; 5--10 Myr), to constrain the form and shape of the luminosity function and mass spectrum in this regime. We conducted a deep multi-band ($Y$\,=\,21.2, $J$\,=\,20.5, $Z$\,=\,22.0 mag) photometric survey of six square degrees in the central region of Upper Scorpius. We extend the current sequence of astrometric and spectroscopic members by about two magnitudes in $Y$ and one magnitude in $J$, reaching potentially T-type free-floating members in the association with predicted masses below 5 Jupiter masses, well into the planetary-mass regime. We extracted a sample of 57 candidates in this area and present infrared spectroscopy confirming two of them as young L-type members with characteristic spectral features of 10 Myr-old brown dwarfs. Among the 57 candidates, we highlight 10 new candidates fainter than the coolest members previously confirmed spectroscopically. We do not see any obvious sign of decrease in the mass spectrum of the association, suggesting that star processes can form substellar objects with masses down to 4--5 Jupiter masses.
\end{abstract}



%
%
\section{Introduction}
\label{USco_VISTA_deepY:intro}

The existence of brown dwarfs and exoplanets is now well established
with several hundreds of examples discovered over the past 25 years
\citep{mayor95,rebolo95,nakajima95}. Large-scale surveys for field ultracool
dwarfs and deep pencil-beam surveys of specific regions have now identified
substellar objects down to a few Jupiter masses with properties differing
depending on their age, environment or metallicity. The shape of the
substellar present-day mass function \citep{salpeter55,miller79,scalo86}
in the solar neighbourhood seems to indicate a power law index $\alpha$ of 0.6, 
where dN/dM\,$\propto$\,M$^{-\alpha}$ accounting for all members of the 20 pc
sample \citep{kirkpatrick19}.
This slope is consistent with the substellar shape of the mass functions
in star-forming regions \citep[review by][]{bastian10}, nearby moving groups
\citep{liu13a,gagne15a,faherty16a}, and young clusters \citep[review by][]{luhman12b}.
How massive can the least massive fragment that star formation mechanisms
can form (also referred to as the limit of fragmentation), 
remains an open question. Early estimates suggested
7 Jupiter masses (hereafter M$_{\rm Jup}$)
\citep{low76,rees76,silk77a} but this limit may change when including
rotation or magnetic fields in the simulations 
\citep{boss88,boyd05,whitworth06,boley10,kratter10b,forgan11,rogers12}.
From the aforementioned studies of nearby substellar objects in the solar vicinity
and in young regions, the limit seems to be consistently below 5 M$_{\rm Jup}$.

Upper Scorpius (USco) is part of the nearest OB association to the Sun,
Scorpius Centaurus. The region is located at 145 pc from the Sun \citep{deBruijne97} and
has an age of 5--10 Myr \citep{preibisch99,preibisch01,pecaut12,song12,david19a}. USco
members show a significant mean proper motion compared to stars along its line of sight
\citep[mean value of $-$11 and $-$25 mas/yr in right ascension and declination, respectively;][]{deBruijne97,deZeeuw99}.
The high-mass population of USco has been explored in X-rays
\citep{walter94,kunkel99,preibisch98}, astrometrically \citep{deBruijne97,deZeeuw99,cook17a,luhman20}
in the optical \citep{ardila00,martin04,slesnick06,slesnick08}, and in the infrared
\citep{lodieu06,lodieu07a,kraus08a,bejar08,lafreniere10a,dawson11,lodieu11c,lafreniere11,dawson13,lodieu13c,lafreniere14,dawson14,best17a}.
Low-mass stars and brown dwarfs in USco have been subject to numerous studied over the
past years thanks to the arrival of wide and deep surveys
\citep{preibisch01,preibisch02,lodieu07a,lodieu11a,penya16a}, and more recently the 
$Gaia$ mission \citep{Gaia_Brown2018},
yielding a relatively well-defined mass function below the stellar/substellar limit in this region \citep{luhman20}.
The Kepler K2 mission \citep{borucki10,lissauer14,batalha14} has revealed the 
first eclipsing binaries in USco over a wide range of masses, giving the first independent
mass and radius determinations at 5--10 Myr \citep{alonso15a,kraus15a,lodieu15c,david16a,david19a}
as well as the first Neptune-size planet orbiting a M3 star \citep{mann16b,david16b}.

\citet{lodieu18a} recently identified the first L dwarf sequence in USco with optical
and infrared spectroscopic characterisation. These authors identified L-type candidates 
in a deep $ZYJ$ survey \citep{lodieu13d} conducted with the Visible Infrared Survey Telescope \citep[VISTA;][]{emerson01,dalton06}. In this paper we release a new, $Y$-band survey 
2 magnitudes deeper over half of the area compared to the former survey 
to look for isolated planetary-mass candidates with ages of 5--10 Myr and investigate the
shape of the luminosity and mass function below the deuterium-burning limit.
In Section \ref{USco_VISTA_deepY:VISTA_surveys} we present the deep infrared observations
obtained with European Southern Observatory (ESO) VISTA/VIRCAM located in Paranal (Chile) 
and the data reduction methods.
In Section \ref{USco_VISTA_deepY:Subaru_survey} we detail the Sloan $z$-band survey conducted
with the Hyper Suprime-Cam camera on Subaru in Mauna Kea (Hawaii, USA).
In Section \ref{USco_VISTA_deepY:selection_cand} we describe the selection of potential planetary-mass member candidates of the USco association. 
In Section \ref{USco_VISTA_deepY:discuss} we discuss the nature of our candidates and
the implications on the shape of the IMF and the theory of the fragmentation limit.

%
%
\section{The VISTA deep surveys}
\label{USco_VISTA_deepY:VISTA_surveys}

This study makes use of two different but complementary VISTA surveys described below. 
VISTA is a 4-m telescope \citep{emerson01,emerson04} based in Paranal (Chile) and 
equipped with the VISTA InfraRed CAMera \citep[VIRCAM;][]{dalton06}. VIRCAM is fitted
with 16 infrared detectors with pixels of 0.339 arcsec offering a field of view of 
1.65 square degrees.

\subsection{VISTA/VIRCAM observations}
\label{USco_VISTA_deepY:VISTA_survey_obs}

On the one hand, we re-processed the deep VISTA $ZYJ$ survey described in \citet{lodieu13d}
to extract PSF photometry instead of aperture photometry for optimal completeness and best accuracy for faint point sources in these relatively crowded fields.
To summarise, these surveys cover 13.5 square degrees towards the central region of USco.
The observations took place between April and May 2012 down to 100\% completeness limits 
of $Z$\,=\,22.0 mag, $Y$\,=\,21.2 mag, 
and $J$\,=\,20.5 mag, respectively. We refer the reader to \citet{lodieu13d} for more details.
We will refer to that survey as the `first epoch' throughout the paper. This work identified
tens of brown dwarf and planetary-mass members, 12 of them being confirmed
spectroscopically as 5--10 Myr-old L1--L7 members \citep{lodieu18a} with an independent classification
by \citet{luhman18a}.

\begin{figure}
  \centering
  \includegraphics[width=\linewidth, angle=0]{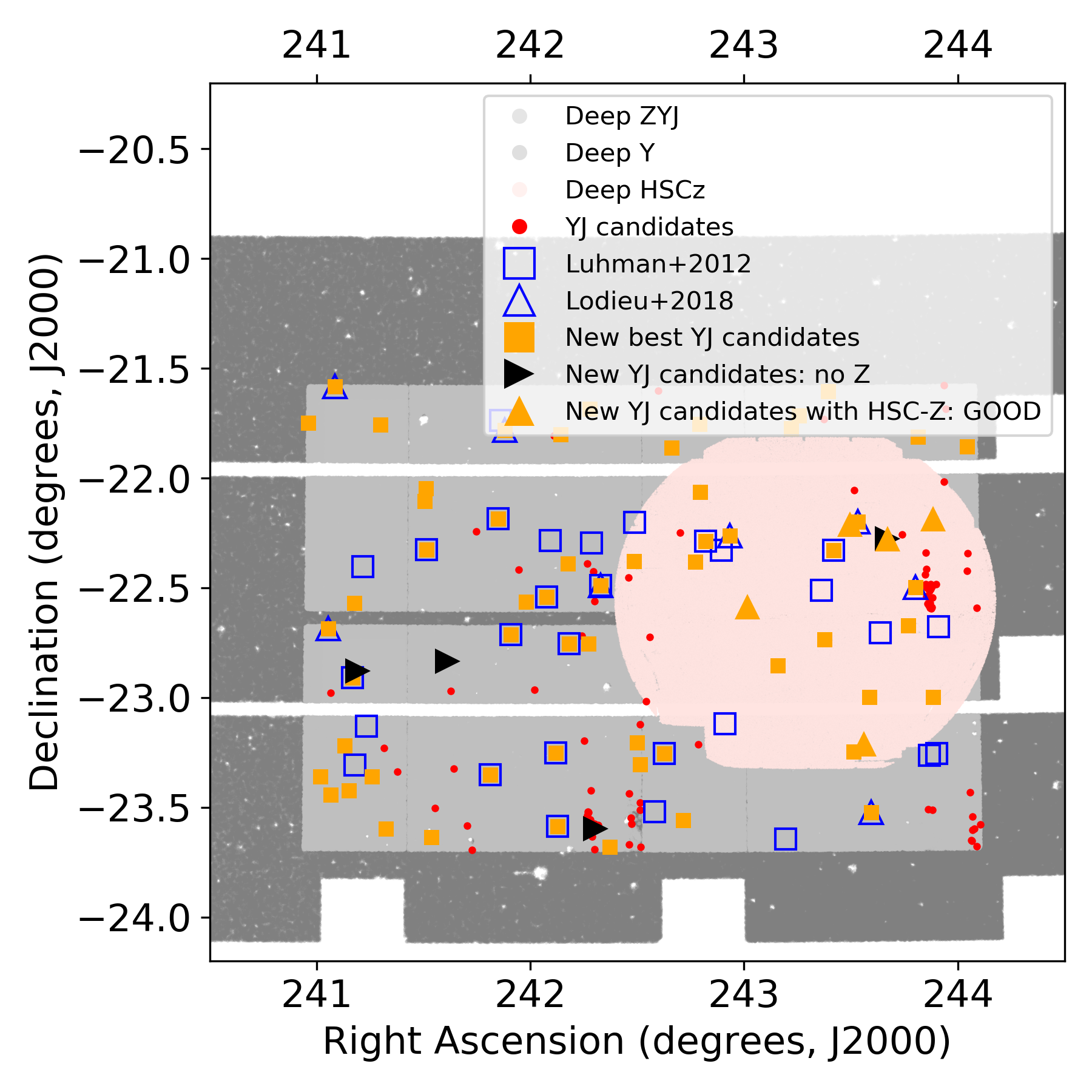}
  \caption{Location of the four deep $Y$-band tiles (light grey) on top of the nine first epoch observations
conducted in $ZYJ$ (grey) with VISTA \citep{lodieu13d}. Overplotted is the coverage of the deep Subaru HSC survey (light rose).
Known members from \citet{luhman12c} 
and \citet{lodieu18a} are shown as blue open squares and triangles, respectively. 
The new candidates selected in this work are highlighted as black and orange symbols.
}
  \label{fig_USco_VISTA_deepY:radec_cover}
\end{figure}

On the other hand, we conducted a deeper $Y$-band survey with VISTA/VIRCAM to be sensitive
to fainter and cooler USco members because the first epoch was limited in depth by the $Y$ and
$Z$ filters. We observed four VISTA/VIRCam tiles in the $Y$-band filter (centered at 1.02$\pm$0.10 micron) between May and July 2016 as part of ESO program 095.C-0781(A) (PI Lodieu). 
We gather the logs of observations in Table \ref{tab_USco_VISTA_deepY:log_obs_VISTA}
compiling seeing, airmass, and ellipticity information for each tile.
We set the on-source integrations to 60\,s repeated twice with six paw-print positions and five 
jitter positions, resulting in a 1h total on-source integration for each tile. We repeated this observing 
block seven times during the May--July 2016 period for the four tiles to achieve a 100\% completeness 
limit of $Y$\,=\,22.6 mag over 6.6 square degrees. Due to the differences in the tiling 
configurations automatically processed by the ESO survey definition tool, the common area between
the two VISTA survey is about six square degrees (Fig.~\ref{fig_USco_VISTA_deepY:radec_cover}).

%
%
\begin{table*}
 \centering
 \caption[]{Logs of the VISTA/VIRCAM $Y$-band observations. We list the name of the tiles
 with their coordinates in sexagesimal format, the range of date$+$time of observations, the total on-source
 exposure time, the median seeing, mean ellipticity, and airmass at the time of observations. We give the average
 parameters for each stacked tile made of seven individual tiles combined together.
 }
{\normalsize
 \begin{tabular}{@{\hspace{0mm}}c c c c c c c c c@{\hspace{0mm}}}
 \hline
 \hline
Tile & R.A.\    &     Dec       &  Date\_min & Date\_max & ExpT & Seeing & Ell & Airmass \cr
 \hline
          & hh:mm:ss.ss & ${^\circ}$:$'$:$''$ & yyyy-mm-dd & yyyy-mm-dd & seconds & arcsec &  &  \cr
 \hline
tile112 & 243.2118900 & $-$23.0915800 & 2016-05-19 06:25:08.3 & 2016-07-12 23:17:48.2 & 8200 & 1.04 & 0.05 & 1.16 \cr
tile111 & 241.6261650 & $-$23.0908100 & 2016-05-18 02:54:48.4 & 2016-06-29 23:57:34.6 & 8200 & 0.92 & 0.05 & 1.18 \cr 
tile122 & 243.2064900 & $-$21.9993800 & 2016-06-01 06:01:13.2 & 2016-07-11 23:31:25.3 & 8200 & 1.03 & 0.05 & 1.13 \cr
tile121 & 241.6331400 & $-$21.9986100 & 2016-06-01 04:49:03.2 & 2016-06-30 05:00:20.3 & 8200 & 1.03 & 0.05 & 1.13 \cr  
\hline
 \label{tab_USco_VISTA_deepY:log_obs_VISTA}
 \end{tabular}
}
\end{table*}
\subsection{Point Spread Function (PSF) Photometry}
\label{USco_VISTA_deepY:VISTA_survey_phot}

VISTA Science Archive (VSA; \citealt{cross12}) standard pipeline products consist of instrumentally 
corrected images combined into stacks and source catalogue extraction with fixed apertures from
those images. While the fixed--aperture source extraction \citep{irwin04} is optimised for faint point 
sources, the detection and deblending of faint stars, in particular those near to much brighter objects, 
is not optimal. Our deepest stacked images are rather more crowded than the shallow and wide 
survey images for which the aperture source extraction has been optimised, so we employed Point 
Spread Function (PSF) fitted photometry in creating our source detection lists from the
VSA image products \citep[see for example][]{mauro13a}.

We employed DAOPHOT~\citep{stetson87} in a recent incarnation running within the 
IRAF/PyRAF environment \citep{tody86,tody93,davis99a,greenfield06a,pyraf12a}.
We followed standard best--practice in detecting sources, defining the PSF and then using that PSF 
to fit all detected objects, simultaneously within small groups, to extract source positions and fluxes. 
Matched--filter source detection at a $2.8\sigma$ threshold was run on all the images using the 
DAOFIND task. A selection of 100 PSF stars was made for each image automatically from the DAOFIND 
output, where every PSF star was required to have a magnitude between 3.0 and 3.5 above the
detection limit (i.e.\ intermediate in brightness for good signal--to--noise but not so bright as to
be affected by non--linearity or saturation), and having no other detected source within two PSF
fitting radii. Additionally we required our PSF stars to have values of the DAOFIND shape parameters
SHARP, SROUND and GROUND within 1$\sigma$ of their median values in order to restrict the selection
to faint, point--like images in every case. The PSF model in DAOPHOT consists of an underlying analytical
function with additional residual corrections \citep{stetson87}. In order to choose the most appropriate analytical model amongst the available options, we made trial fits and compared the residual profile 
scatter for all. We finally settled on the `penny2' option (a Gaussian core with Lorentzian wings, both of
which are elliptical, and tilted arbitrarily and independently to the image coordinate axes). In order to 
account for position--dependent variation in the PSF we selected `varorder' equal to~2 for quadratically 
varying terms in the residual correction look--up tables of the combined PSF model. A default PSF 
model radius of 11~pixels was employed during PSF modelling, while a PSF fitting radius of twice 
the full-width-half-maximum (FWHM) measured in each image was employed at the PSF photometry stage when photometring 
all detected sources.

We matched the $Y$ and $J$ catalogues within a conservative radius of 3~arcsec, yielding
a total of 887,175 sources. We applied two quality criteria on the sharpness (between $-$0.5
and 0.5) and $\chi^2$ (less than 3.0) parameters in both filters that returned 617,715 objects, 
which will be our input catalogue in the rest of our analysis. 
We note that the number of sources with pairing radii larger than 1 arcsec is relative constant.
We count 7106 pairs beyond 1 arcsec, resulting in a contamination less than 1.2\%.
We show the resulting ($Y-J$,$J$) colour-magnitude diagram in the left-hand side
of Fig.\ \ref{fig_USco_VISTA_deepY:CMD_YJY}. In the right-hand side of 
Fig.\ \ref{fig_USco_VISTA_deepY:CMD_YJY}, we plot the same diagram and overplot known 
USco members from the literature and our photometric candidates.

We cross-matched the $Y,J$ catalogue with photometry from our previous VISTA $Z$-band 
as well as $H+K$ from the UKIRT Infrared Deep Sky Survey \citep[UKIDSS;][]{lawrence07} 
Galactic Clusters Survey (GCS) Data Release 9 (DR9), the AllWISE survey \citep{cutri13,cutri14}, 
and the  Panoramic Survey Telescope and Rapid Response System catalogue \citep[Pan-STARRS;][]{chambers16a} with a pairing radius of 3 arcsec in each case.
We will make the full catalogue available to the community via Vizier at
the Centre de Donn\'ees de Strasbourg. The full table contains the coordinates in J2000, 
the magnitudes in each filter with their associated uncertainties, the sharpness and
$\chi^2$ parameters given by {\tt{daophot}} and {\tt{allstar}} in IRAF, the
proper motion in mas, and the cross-match with other catalogues like the UKIDSS GCS DR9,
AllWISE, and PanStarrs DR1\@.

%
%
%

%
%
\section{The Hyper Suprime-Cam survey}
\label{USco_VISTA_deepY:Subaru_survey}
\subsection{Subaru Hyper Suprime-Cam observations}
\label{USco_VISTA_deepY:Subaru_survey_obs}

The Hyper Suprime-Cam (HSC) is a very large charge--coupled detector (CCD) mosaic camera mounted at 
prime focus of the Subaru 8.2-m Telescope \citep{iye04a} in Mauna Kea, Hawaii. The HSC uses 104 science charge-coupled devices sensitive to optical wavelengths, offering a 1.5-degree field-of-view in diameter with a pixel size of 0.17 arcsec \citep{miyazaki18a,komiyama18a,kawanomoto18a,furusawa18a}.
The auto-guiding is ensured with four CCDs while the focus is monitored with eight CCDs.
We employed the Sloan $z$-band filter centered at 8921.66 \AA{} with a width of 792.99 \AA{} to be as sensitive as possible to the
coolest members of the association, with the goal of improving on the depth of the deep $Z$ survey
(Section \ref{USco_VISTA_deepY:VISTA_survey_obs}).

The observations were collected in service mode on 25 May 2020 between UT\,=\,10h58 and 12h40 as part of programme S20A-QN098 
(PI Lodieu). The airmass of USco from Mauna Kea was between 1.4 and 1.8\@. The seeing was better than 0.8 arcsec. 
The transparency was higher than 90\% and the sky was dark with no moon.
We collected four blocks of six integrations of 252 seconds, yielding a total 
exposure of 5040s on a single 1.5-degree field centered 
at RA\,=\,16h13m, dec\,=\,22$^{\circ}$36$'$ (Fig.\ \ref{fig_USco_VISTA_deepY:radec_cover}).

\subsection{Data reduction and catalogue generation}
\label{USco_VISTA_deepY:Subaru_survey_phot}

A standard data reduction procedure was employed in processing the HSC via the LSST pipeline stack \citep{jenness2017}.
Instrumental correction consisted of bias prescan/overscan, dark, illumination and flat corrections along with a default
brighter--fatter kernel correction \citep{bosch2018}. Single--visit processing to create accurate astrometric and
photometric calibration at the individual CCD level was performed with respect to the PanSTARRS DR1 catalogue \citep{chambers16a}. 
The 20 individual frames from each of the 104 CCDs available within the field visit where warped and stacked in patches 
to create a deep stack image from which sources were extracted and measured for point-spread function (PSF) fitted photometry, 
again using the standard LSST routines. The HSC--$Z$ band photometry, by default on an AB magnitude scale, was transformed to 
the VIRCam Vega system using a correction $z_{\rm Vega}\,=\,z_{\rm AB} - 0.521$ \citep{cross12}. Finally, a small $\pm0.05$ 
linear colour correction was applied to the photometry as a function of $z-J$ colour to put the HSC--$Z$ measurements on the 
natural VIRCam $Z$ photometric system.

We extracted a total 853,676 sources with a completeness limit of $Z$\,=\,24.3 mag, reduced to 379,279 objects when removing saturated sources (typically brighter than 17.5 mag), and extended sources as well as those with poor PSF fitting.
We cross-matched the full deep $ZYJ$ VISTA catalogue with the good-quality sources in the HSC catalogue and found 162,735 objects in common.
We will make this full VISTA$+$Subaru catalogue of identifiers, coordinates, $Z$ photometry with its associated error, classification (point source or extended), and flag available 
to the community via Vizier.

%
%
\section{Identification of faint USco member candidates}
\label{USco_VISTA_deepY:selection_cand}

The goal of this section is to focus on member candidates fainter than the coolest L dwarfs 
identified in our original $ZYJ$ survey \citep{lodieu13d} and recently confirmed spectroscopically 
\citep{lodieu18a} by combining the deep $Y$-band second epoch survey with the first epoch $ZYJ$ survey.
To set the scene, the two L7 spectroscopic members have $Y$\,=\,20.8--21.0 mag, $J$\,$\sim$\,19.3 mag,
and $Z$ magnitudes beyond the 100\% completeness limit of the first epoch ($Z$\,=\,22 mag).
We start by reviewing our current knowledge on the photometric properties of field and
young L and T dwarfs before jumping into the selection of USco brown dwarfs and isolated
planetary-mass objects.

We emphasise that studies of isolated planetary-mass objects of T-type in clusters and 
star-forming regions younger than 20 Myr 
currently are rather limited. Most objects are candidates whose membership remains under debate
or that need to be confirmed through spectroscopy and/or astrometry 
\citep[e.g.][]{burgess09,haisch10,penya11a,barsony12,spezzi12b,scholz12b,lodieu13b,chiang15b,penya15}.

\subsection{Photometric properties of field L/T dwarfs}
\label{USco_VISTA_deepY:properties_fieldLT}

\begin{figure*}
  \centering
  \includegraphics[width=0.48\linewidth, angle=0]{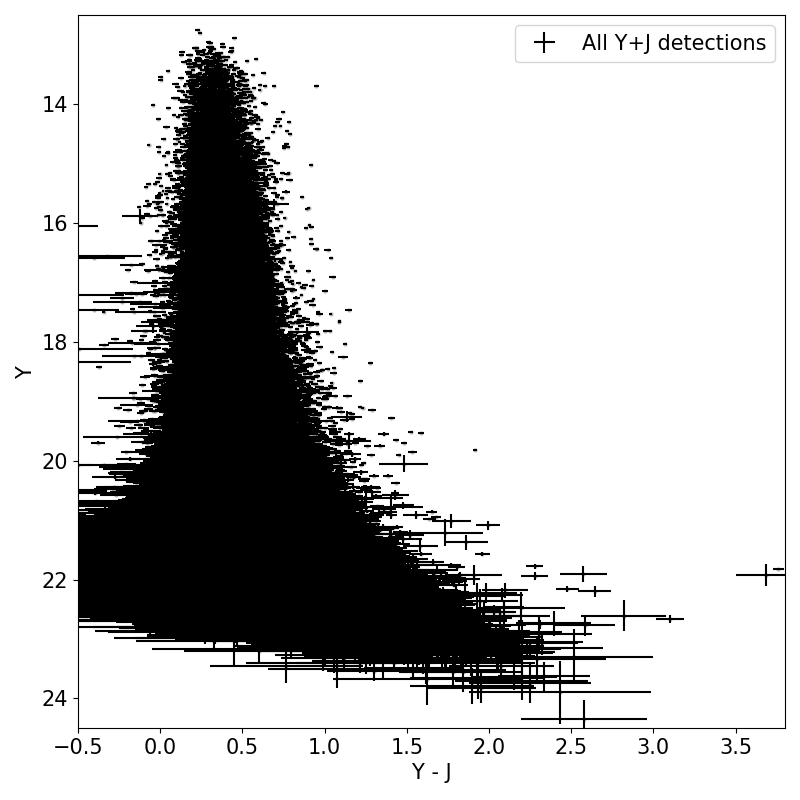}
  \includegraphics[width=0.48\linewidth, angle=0]{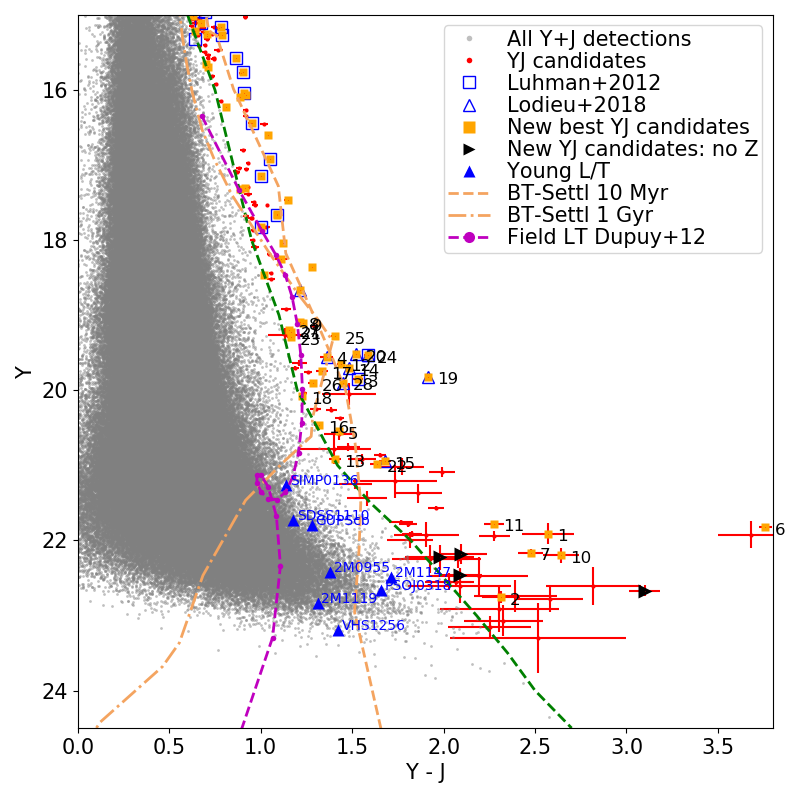}
  \caption{
  {\it{Left:}} ($Y-J$,$Y$) colour-magnitude diagram for all $Y,J$ sources in the two VISTA surveys with their error bars.
  {\it{Right:}} Same as left but with the addition of known USco members (blue open symbols) from \citet{luhman12c} and \citet{lodieu18a}.
  The thick green line depicts our $YJ$ photometric selection criterion.
  Red symbols highlight the new $YJ$ candidates after applying our photometric selection
  (Section \ref{USco_VISTA_deepY:selection_cand_YJ}),
  whereas the orange symbols and black triangles show our best candidates that deserve spectroscopic follow-up. 
  Their ID numbers from the first column of the tables in Appendix \ref{USco_VISTA_deepY:Appendix_Tables} are also indicated.
  The orange dashed lines represent the BT-Settl models for ages of 10 Myr dashed) and 1 Gyr (dot-dashed).
  The purple line depicts the sequence of field L and T dwarfs from \citet{dupuy12}.
  Filled blue triangles and squares show known L and T dwarf members of young moving groups and the Pleiades, respectively.
  }
  \label{fig_USco_VISTA_deepY:CMD_YJY}
\end{figure*}

\begin{figure*}
  \centering
  \includegraphics[width=0.48\linewidth, angle=0]{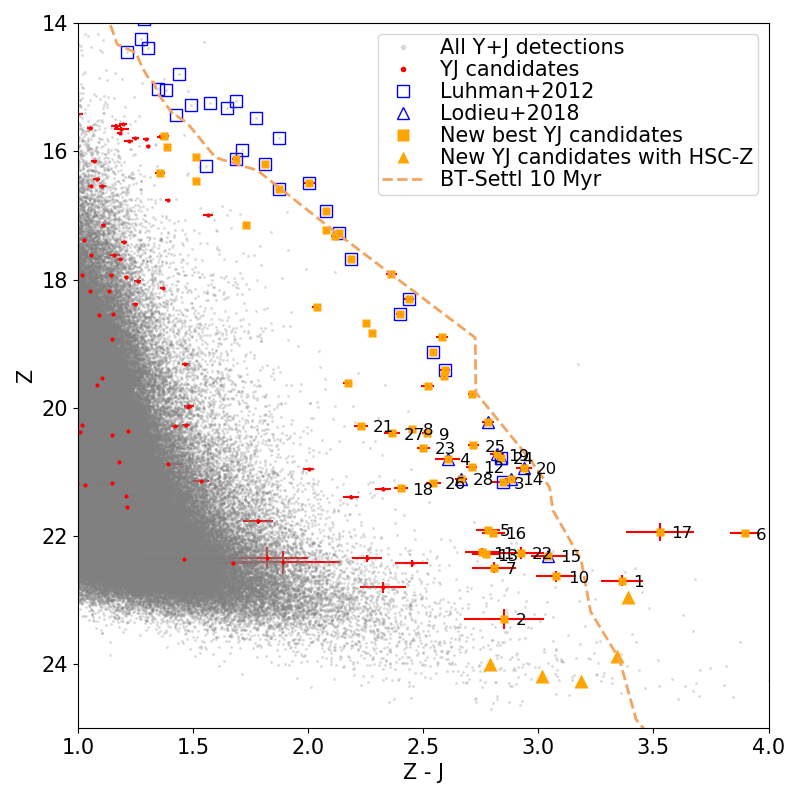}
  \includegraphics[width=0.48\linewidth, angle=0]{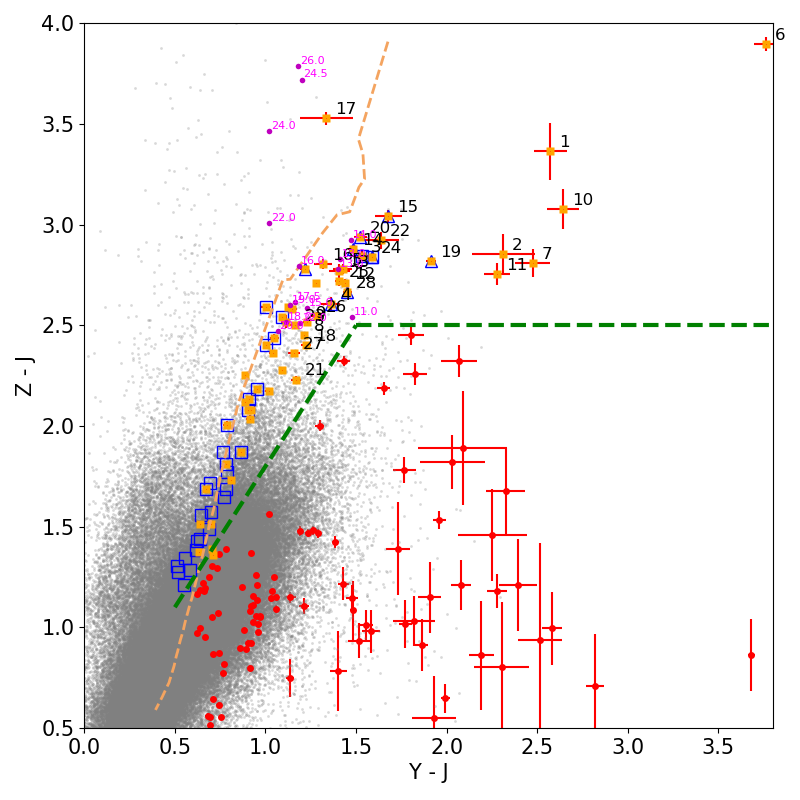}
  \caption{
  {\it{Left:}} ($Z-J$,$Z$) colour-magnitude diagram for $YJ$ candidates with $Z$-band detections (grey dots).
  Blue open squares and triangles are known USco members from \citet{luhman12c} and \citet{lodieu18a}. The dashed brown line depicts the 10 Myr-old
  isochrone from \citet{baraffe15}.
  Red dots highlight USco member candidates after applying the selection in the ($Y-J$,$Y$) diagram.
  Orange and black symbols highlight our best L/T transition candidates for future spectroscopic follow-up.
  {\it{Right:}} ($Y-J$,$Z-J$ colour-colour diagram with the same legend of symbols.)
  }
  \label{fig_USco_VISTA_deepY:CMD_ZJZ}
\end{figure*}

\begin{figure*}
  \centering
  \includegraphics[width=0.49\linewidth, angle=0]{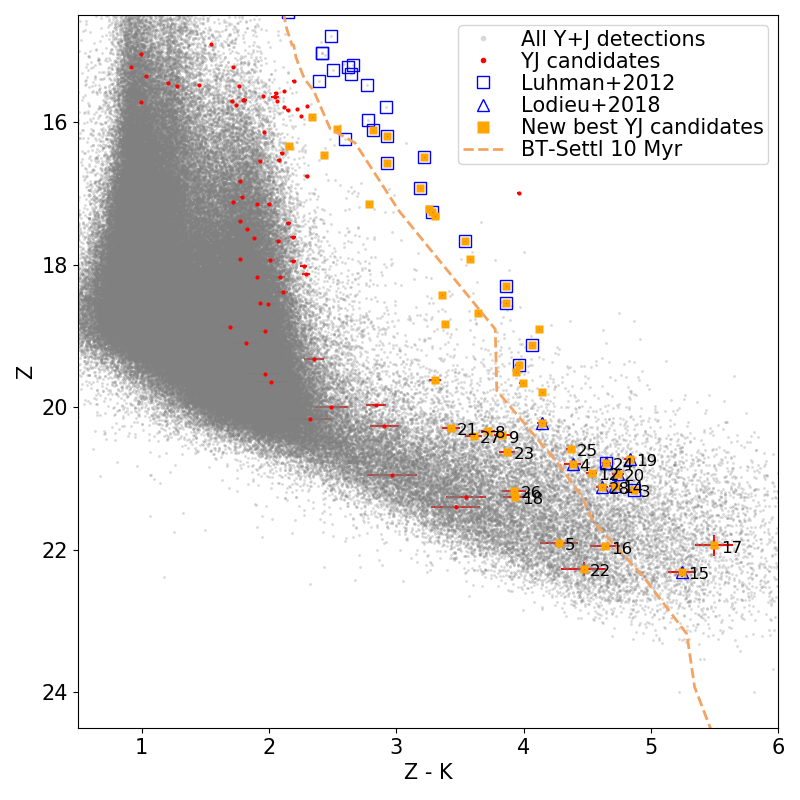}
  \includegraphics[width=0.49\linewidth, angle=0]{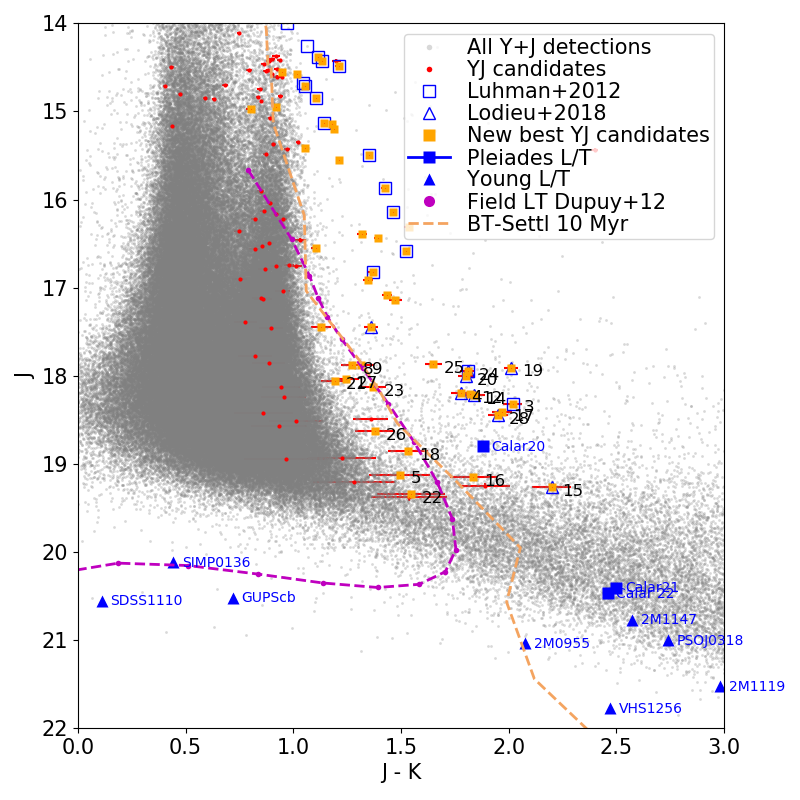}
  \caption{($Z-K$,$Z$) and ($J-K$,$J$) colour-magnitude diagrams for $YJ$ candidates with 
  $Z$-band photometry and $K$-band counterparts from the UKIDSS GCS DR10\@.
  Red squares and blue triangles are known USco members from
  \citet{luhman12c} and \citet{lodieu18a}.
  Orange symbols and black triangles highlight our final candidates after applying filtering in the $YJ$, $ZJ$, and $ZK$ colours.
  The purple line with dots represents the sequence of field M, L, and T dwarfs \citep{dupuy12}.
  The filled blue triangles show three L7--L8 members of TWA \citep{gagne17a} and SIMP\,J013656.5+093347
  a member of the 200 Myr-old Carina-Near moving group \citep{gagne17b}.
  The filled blue squares mark the position of the Pleiades member candidates Calar 20, 21, and 22
  \citep{zapatero14c}.
}
  \label{fig_USco_VISTA_deepY:CMD_ZKZ_JKJ}
\end{figure*}

We looked at the colours and absolute magnitudes of field L and T dwarfs 
\citep{dupuy12} and located them in several diagrams (see figures in 
this paper) to guide the identification of their younger counterparts. 
We draw the following conclusions, keeping in mind that our current poor 
knowledge of the spectral energy distribution of T-type objects at young ages.
\begin{itemize}
\item The $Y-J$ colours of field L/T dwarfs are approximately constant around 1 mag, going from 1.2 mag 
for L6 to 1.1 mag to T8 but with values slightly below 1.0 mag for T2 and T3 types \citep{dupuy12}. From the synthetic colours of L/T dwarfs \citep{hewett06}, a kink towards the blue is observed in the $Y-J$ colours for transition sources followed by a relatively constant value from mid-T dwarfs
(right-hand side panel in Fig.\ \ref{fig_USco_VISTA_deepY:CMD_YJY}).
\item The $Z-J$ vs $Y-J$ diagram indicates that L/T transition objects have $Z-J$ colours
as blue as 2.5 mag \citep{hewett06} with a larger dispersion for L/T transition dwarfs 
and an average $Z-J$ colour that keeps being redder with later spectral types 
(Fig.\ \ref{fig_USco_VISTA_deepY:CMD_ZJZ}).
\item The $Z-K$ colours of L/T transition objects show a similar dispersion as in $Z-J$
while their $J-K$ colours become bluer from T0 ($J-K$\,=\,1.4 mag) to T1 ($J-K$\,=\,1.13 mag)
as seen in the right panel in Fig.\ \ref{fig_USco_VISTA_deepY:CMD_ZKZ_JKJ}.
\item The absolute $Y$ magnitude of a T4 is as bright as a L7 but slightly fainter than a L6\@.
\item A old field T4 dwarf has SDSS$z$\,$\sim$\,23.8 mag \citep{hewett06,schmidt10b} and 
$Y,J$\,=\,21.13, 20.13 mag \citep{dupuy12} but USco members tend to be 1 mag brighter than field objects.
\end{itemize}
\subsection{Photometric properties of young L/T dwarfs}
\label{USco_VISTA_deepY:properties_youngLT}

We examined the colours of five red young L7 dwarfs within 50 pc with ages
less than 20 Myr (determined via membership of young moving groups) similar to USco
recently published in the literature
\citep[e.g.][]{looper07,gauza15a,gagne15c,filippazzo15,faherty16a,liu16a,schneider16b}.
We also compiled three T dwarfs with older ages but well-constrained ages based on their 
high-probability membership of
the AB\,Dor \citep[GUPSc\,b (T3.5), SDSS\,J11101001$+$0116130 (T5.5); 70--130 Myr;][]{gagne17a} and 
Carina-Near \citep[SIMP01365662$+$0933473 (T2.5); 150--300 Myr;][]{gagne17b} young moving groups
plotted as blue filled symbols in all colour-magnitude diagrams of this manuscript.
Unfortunately, there are no $Z$ magnitudes published for these sources, which appears as
a key filter in our study.
We also added to the diagrams three L/T transition member candidates (blue squares) 
of the Pleiades \citep{zapatero14b} at an age of 125 Myr \citep{stauffer98} identified 
in a deep photometric survey with astrometric information \citep{zapatero14a}.
We plot all these young L/T transition sources with $Y,J,w1,w2$ photometry compiled from the literature in the diagrams presented in this manuscript
after correcting their apparent magnitudes for their distances and applying
the distance modulus of the USco association \citep[145 pc;][]{luhman20a}.
We note that all five young L7 dwarfs are intrinsically fainter than the USco L7 members. They show
bluer $Y-J$ colours but redder colours in other combinations of filters that may come from
differences in the optical thickness of the dust cloud deck \citep{marocco14}.

%
%
%

%
%
%
%

%
%
%

%
%
%
\begin{figure*}
  \centering
  \includegraphics[width=0.49\linewidth, angle=0]{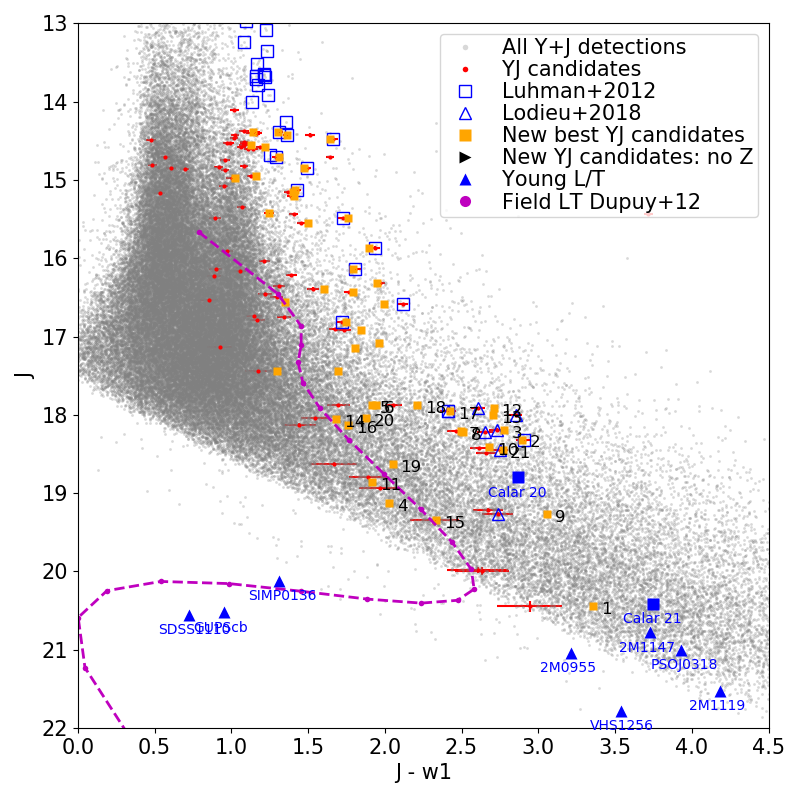}
  \includegraphics[width=0.49\linewidth, angle=0]{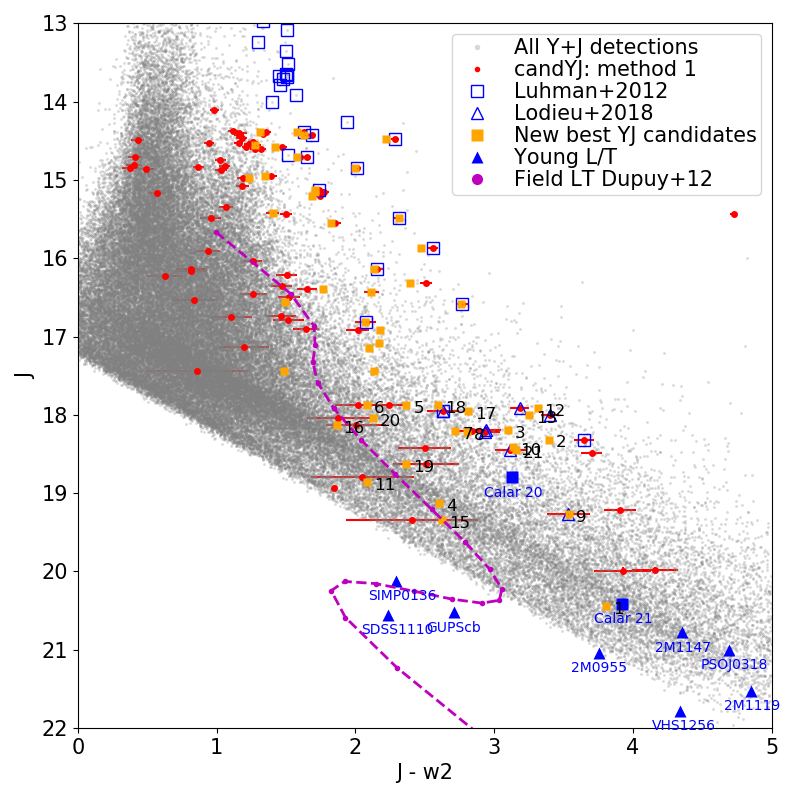}
  \includegraphics[width=0.49\linewidth, angle=0]{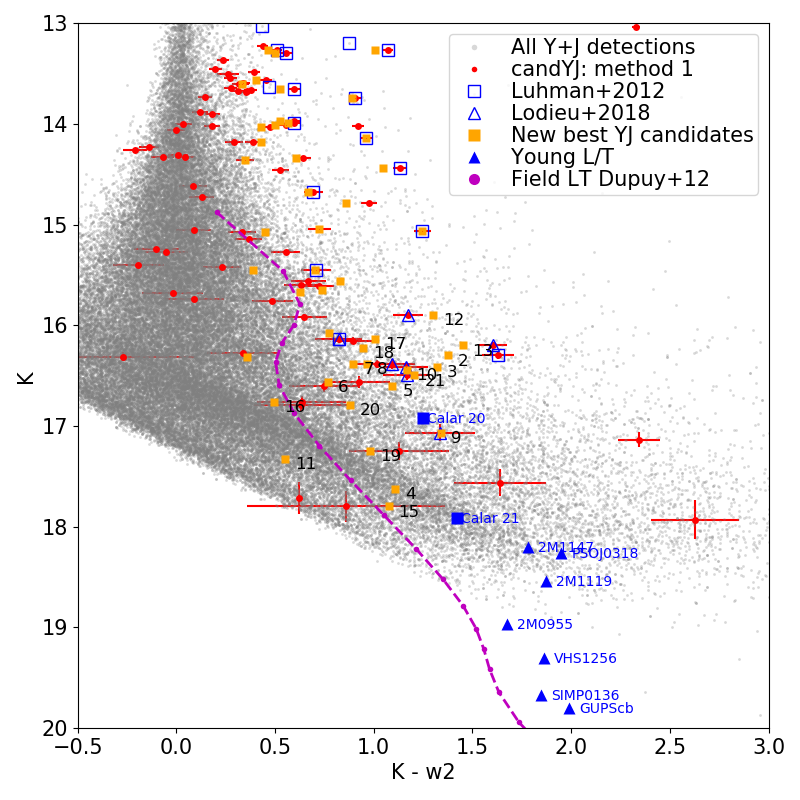}
  \includegraphics[width=0.49\linewidth, angle=0]{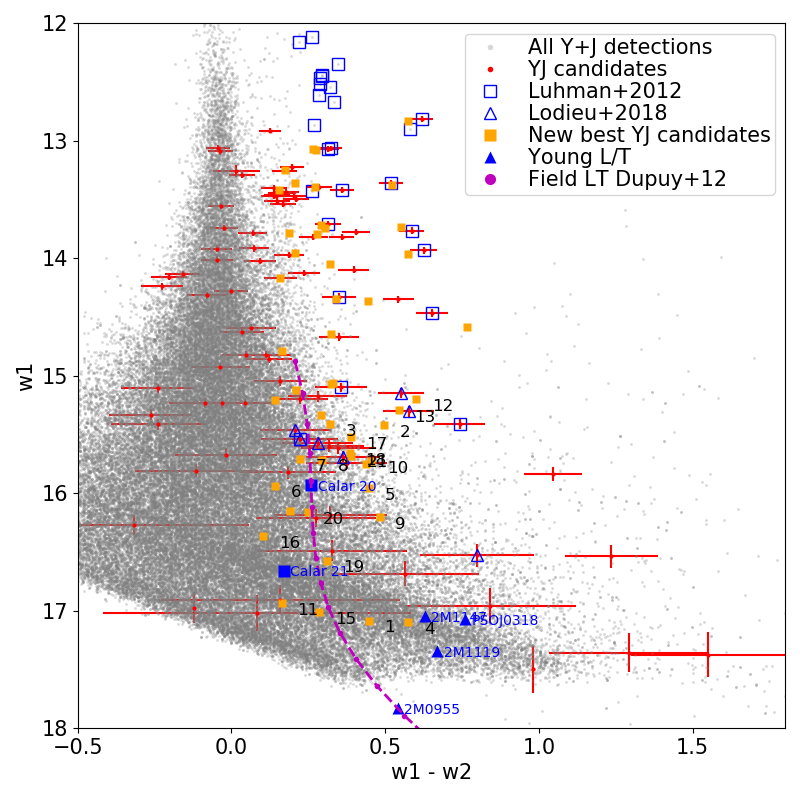}
  \caption{Colour-magnitude diagrams displaying all $YJ$ candidates with $Z$-band photometry
  and UKIDSS GCS DR10 and AllWISE/unWISE counterparts.
  Red squares and blue triangles are known USco members from \citet{luhman12c} and \citet{lodieu18a}.
  Orange symbols highlight our final candidates after applying filtering in the $YJ$, $ZJ$, and $ZK$ colours.
  Orange symbols and black triangles highlight our best candidates with unWISE photometry.
  The purple line with dots represents the sequence of field M, L, and T dwarfs \citep{dupuy12}.
  The blue filled triangles show three L7--L8 members of TWA \citep{gagne17a} and SIMP\,J013656.5+093347
  a member of the 200 Myr-old Carina-Near moving group \citep{gagne17b} while the blue filled squares
  mark the position of the Pleiades member candidates Calar 20, 21, and 22
  \citep{zapatero14c}.
}
  \label{fig_USco_VISTA_deepY:CMD_WISE}
\end{figure*}
\subsection{YJ photometric candidates}
\label{USco_VISTA_deepY:selection_cand_YJ}

As a first step, we carefully looked at the ($Y-J$,$Y$) diagram in the left-hand side panel
in Fig.\ \ref{fig_USco_VISTA_deepY:CMD_YJY}. We suggest that the USco sequence is visible, 
running from around ($Y-J$,$Y$) = (0.75, 15.0) to (1.45, 20.0) to (2.6, 22.5),
with the reddest candidate at around (3.1, 22.7). 
The first part of the USco sequence down to $Y$\,=\,20 mag is supported by known
members while the bottom part is based on the morphology of the \CMD{}.
To confirm this hypothesis, we overplot on this diagram 
known USco members from \citet{luhman12c} and
spectroscopic L dwarf members \citep{lodieu18a}, displayed as open blue 
squares and triangles in Fig.\ \ref{fig_USco_VISTA_deepY:CMD_YJY}, respectively.
Hence, we designed a series of selection lines following the USco sequence but shifted
to the blue of the sequence to be conservative. We limited our search to sources 
fainter than $Y$\,=\,15 mag because brighter sources tend to be saturated in our deep 
survey. Moreover, our goal is to identify the coolest L and T-type members and 
this region has been well surveyed at brighter magnitudes by various teams
\citep[e.g.][]{lodieu07a,dawson11,lodieu11c,luhman12c,dawson13,best17a}.
This selection returned 157 candidates and is abbreviated below as ``New cand$YJ$''
(red dots in all figures). Among these 157, 21 and 16 have matching radii larger than
1.0 and 1.5 arcsec, respectively, yielding a level of contamination of about 10\% due to our
conservative pairing.
We now proceed to further constrain their membership
based on other colour-magnitude diagrams.

\subsection{Red-optical Z-band photometry}
\label{USco_VISTA_deepY:selection_cand_Z}

We investigated the position of the $YJ$ candidates in the ($Z-J$,$Z$) diagram
to further assess their membership (left-hand side panel in Fig.\ \ref{fig_USco_VISTA_deepY:CMD_ZJZ}). 
We found 153 sources (out of 157) with $Z$-band photometry from the first epoch of our 
original VISTA $ZYJ$ survey, the remaining four are $Z$ dropouts. 

Fig.\ \ref{fig_USco_VISTA_deepY:CMD_ZJZ} depicts the ($Z-J$,$Z$) diagram with
the same symbols as the ($Y-J$,$Y$) diagram.
We observe that the sequence of members is well defined and becomes redder down 
to $Z$\,$\sim$\,20 mag, as confirmed by the compilation of members of \citet*{luhman12c}
and highlighted in our previous studies \citep[e.g.][]{lodieu06,lodieu07a,lodieu13c}. 
Beyond, the sequence seems to flatten with a relatively constant $Z-J$ colour and
an increased dispersion that might be due to a combination of physical spread
and large error bars at those faint magnitudes or intrinsic physical properties
of such objects (dust, cloud decks).

We show the ($Y-J$,$Z-J$) colour-colour diagram in the right panel of
Fig.\ \ref{fig_USco_VISTA_deepY:CMD_ZJZ}. Based on known members, we can
see the USco sequence. We designed two straight lines to discard photometric
non-members based on these two colours (green dashed lines in the right 
panel of Fig.\ \ref{fig_USco_VISTA_deepY:CMD_ZJZ}): the first one goes from
($Y-J$,$Z-J$)\,=\,(0.5,1.1) to (1.5,2.5) mag while the second one excludes all
sources with $Z-J$ colours bluer than 2.5 mag. The former is driven by the
position of bright USco members from \citet*{luhman12c} while
the limit on the $Z-J$ colour corresponds to the
reddest field M dwarfs in the solar neighbourhood \citep{schmidt10b}.
These selection criteria led to 57 bona-fide members highlighted with
orange squares in all figures in this manuscript
(Table \ref{tab_USco_VISTA_deepY:new_candYJ_best}).

In \citet{lodieu18a} we presented follow-up optical imaging in the Sloan $i$ filter 
on the OSIRIS instrument \citep[Optical System for Imaging and low Resolution Integrated Spectroscopy;][]{cepa00}
installed on the Gran Telescopio de Canarias (GTC) in La Palma (Canary Islands).
The observations were taken as part of programme GTC4-14A during the month of 
August 2014 over several days under variable conditions resulting in 
inhomogeneous depths (Fig.\ \ref{fig_USco_VISTA_deepY:radec_cover}).
We refer the reader to that paper for the logs of the observations and the description 
of the data reduction and analysis. We found 10 $YJ$ candidates covered by the 
GTC observations, nine previously published in \citet{lodieu18b} and one new.
This yet-unpublished SDSS$i$ photometry of VISTA\,J16125313$-$2146274 further confirms
this source as a member with a $i-J$ colour of 2.36$\pm$0.02 mag 
(SDSS~$i$\,=\,17.337$\pm$0.003 mag; $J$\,=\,14.977$\pm$0.020 mag),
in agreement with the expected sequence of USco members.

Among the four (157$-$153\,=\,4) $Z$ dropouts listed in 
Table \ref{tab_USco_VISTA_deepY:new_candYJ_noZ} in Appendix \ref{USco_VISTA_deepY:Appendix_Tables},  none have additional photometric information 
from large-scale surveys such as UKIDSS GCS DR9, AllWISE or PanStarrs. We have checked
the images of these databases and confirm that those positions are void of sources
down to their detection limits. None of these four candidates appear in the unWISE 
catalogue of \citet{schlafly19}.
Their $Y$ and $J$ magnitudes range from 22.18--22.67 mag and 19.57--20.37 mag,
respectively. Except for the source covered by the Subaru survey, we cannot further 
constrain the membership of these sources so 
they remain possible candidates. Nonetheless, we have checked the deep $Y$ images
as well the first VISTA $Z,J$ epoch and can say the following about each object:

\begin{itemize}
    \item 16:09:13.79 $-$23:35:46.9 ($J$\,=\,20.25) lies in the halo of a bright 
    star and might not be a real detection. We cast doubt on the membership of this source to USco.
    \item 16:04:46.7 $-$22:52:41.4 ($J$\,=\,20.37) looks double in the $J$ images
    while the $Y$ image shows a single source. The object remains undetected in $Z$
    and in the UKIDSS images.
    \item 16:06:26.99 $-$22:50:00.6 ($J$\,=\,20.09) lies in the halo of a bright star
    but the detections in $Y$ and $J$ are clear. The object is undetected in $Z$
    and in the images of the UKIDSS GCS\@. This object remains as a good candidate.
    \item 16:14:41.41 $-$22:16:29.8 ($J$\,=\,19.57) looks double in the $J$ images
    while the $Y$ image shows a single source. This is a similar case as the candidate
    in the second bullet above (16:04:46.7 $-$22:52:41.4).
    The object is undetected in the in UKIDSS GCS but is detected as a point source
    in the Subaru $Z$ image 
    with $Z$\,=\,22.959$\pm$0.025 mag. Its position in the ($Y-J$,$Z-J$) and ($Z-J$,$Z$)
    diagrams adds credibility to its membership to the USco association.
\end{itemize}

Up to now, we have assumed that the USco sequence is clearly visible in the
($Y-J$,$Y$) diagram and extend previous compilation of astrometric and spectroscopic members.
To take into account the blue $Y-J$ colours of 
field (Section \ref{USco_VISTA_deepY:properties_fieldLT}) and 
young (Section \ref{USco_VISTA_deepY:properties_youngLT}) L/T transition
brown dwarfs and the extra depth provided by the Subaru HSC survey, we also tried 
applying a very conservative photometric selection, keeping all sources satisfying 
$Y-J$\,$\geq$\,0.9 mag and $Y$\,$\geq$\,19.5 mag undetected in the first VISTA 
$Z$-band epoch. This selection returned 4248 sources, of which 899 have an entry
in the HSC catalogue but only 404 have good quality photometry or are unsaturated. 
Placing these sources in the ($Y-J$,$Z-J$) diagram, we
observed that five remain as potential candidates because of their positions in 
the diagrams involving $ZYJ$ photometry (Table \ref{tab_USco_VISTA_deepY:new_candYJ_HSCZ}
in Appendix \ref{USco_VISTA_deepY:Appendix_Tables}). However, only two of these five 
(the two brightest in $Z$) appear
as most likely members because they extend the current sequence of members in the 
($Z-J$,$Z$) and ($Y-J$,$Z-J$) diagrams (Fig.\ \ref{fig_USco_VISTA_deepY:CMD_ZJZ}).
We emphasise that the HSC survey covers only 1.5 square degrees and that we
recover only $\sim$20\% (899 out of 4248) of the sources without VISTA $Z$ photometry.
Hence, the number of potential L/T transition members in USco might be five times larger,
fact that we should consider when discussing the luminosity function
(Section \ref{USco_VISTA_deepY:discuss_IMF}).

\subsection{Infrared photometry}
\label{USco_VISTA_deepY:selection_cand_NIR}

The $YJ$ candidates with $Z$-band photometry not rejected in the ($Y-J$,$Z-J$) diagram, 
which turns out to be one of the best diagrams to discriminate members from contaminants, 
remain as potential members in diagrams involving near- and mid-infrared photometry 
from the UKIDSS GCS DR9, AllWISE, and unWISE catalogues, respectively 
(Fig.\ \ref{fig_USco_VISTA_deepY:CMD_ZKZ_JKJ} and Fig.\ \ref{fig_USco_VISTA_deepY:CMD_WISE}).
However, we point out that eight sources with the following ID might be non members
based on their positions in several of these diagrams with infrared photometry:
\#5, 8, 9, 18, 21, 22, 23, 26, and 27\@. However, we can not entirely discard them 
as potential members because L/T field transition dwarfs show a significant dispersion
in the $Y-J$ and $Z-J$ colours that we observe in this part of the sequence
(Fig.\ \ref{fig_USco_VISTA_deepY:CMD_ZKZ_JKJ}).

As mentioned in the previous Section \ref{USco_VISTA_deepY:selection_cand_Z}, only one
of the four $YJ$ candidates without VISTA $Z$ detections has additional photometry from 
our Subaru deep survey and remains as bona-fide candidate. The other three sources
do not appear in the infrared CMDs 
(Fig.\ \ref{fig_USco_VISTA_deepY:CMD_ZKZ_JKJ}--\ref{fig_USco_VISTA_deepY:CMD_WISE}).

Finally, as mentioned earlier, the depth of the $K$ and $w1,w2$ surveys are generally insufficient
to detect counterparts to our faintest candidates, especially those with $Z$-band photometry
from the deep HSC survey, and allow for a fair comparison with 
the locations of the young T-type brown dwarfs in colour-magnitude diagrams.
Nonetheless, we find two of the 10 HSC $Z$ detections (but VISTA $Z$ dropouts)
in AllWISE, adding credit to their membership because they follow the empirical
USCo sequence (Fig.\ \ref{fig_USco_VISTA_deepY:CMD_WISE}).

%
%
\begin{figure}
  \centering
  \includegraphics[width=\linewidth, angle=0]{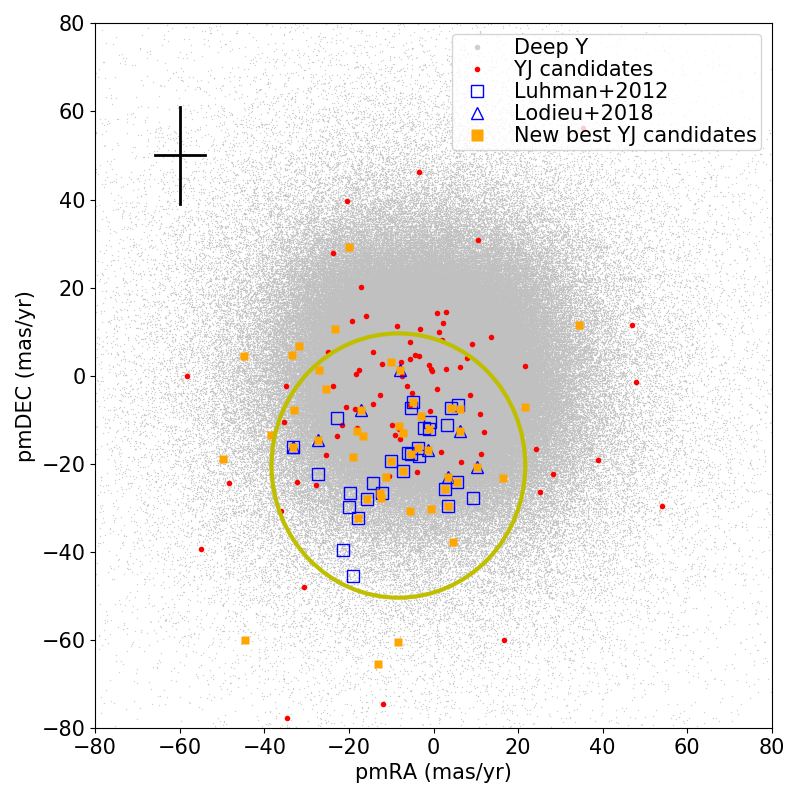}
  \caption{Vector point diagram with the proper motions in the right ascension and declination 
  (in  mas/yr) of the $YJ$ candidates (red dots). Typical error bars are drawn in 
  the top left corner. Blue squares and triangles are known USco members from the compilation
  of \citet{luhman12c} and \citet{lodieu18a}.
  Orange symbols highlight our final candidates after applying filtering in the $YJ$, $ZJ$, and $ZK$ colours.
}
  \label{fig_USco_VISTA_deepY:VPD_diagram}
\end{figure}
\subsection{Proper motions}
\label{USco_VISTA_deepY:selection_cand_PM}

We calculated the proper motion in right ascension and declination of the $YJ$ candidates
comparing the positions between the $J$-band first epoch and the $Y$-band second epoch
assuming a mean baseline of 4.15 years.
We ignore the $Z$ and $Y$ first epochs because they are shallower and do not include a
detection for all the $YJ$ candidates. We calculated the mean dispersion of the full sample
of several hundred thousands $Y$ and $J$ detections in both directions to estimate the error bars 
on the proper motions of the faintest candidates in our sample. We find median absolute deviation 
values below 10.2 and 9.5 mas/yr for sources brighter than $Y$\,=\,21 mag in right ascension and
declination, respectively. The deviations increase to values of about 15 mas/yr , 30 mas/yr, 
and 135 mas/yr in both directions in the $Y$\,=\,21--22, 22--23, 23--24 mag intervals,
respectively, and represent upper limits.
We note that the mean values of the proper motions are $-$1.90 and 0.33 mas/yr in right ascension 
and declination, respectively, hence very close to the expected nominal centre at (0,0).

We cross-matched the list of 863 members from \citet{luhman12c} with the full VISTA catalogue to
estimate the mean proper motion of USco members from the 4.15 year baseline between
the two VISTA epochs, yielding ($\mu_{\alpha}\cos\delta$, $\mu_{\delta}$)\,=\,($-$8.3, $-$20.4) mas/yr
based on 29 sources in common to the area surveyed by our deep $Y$ survey.
These mean values agree well with the mean values of the Hipparcos \citep{deBruijne97,deZeeuw99}
and $Gaia$ \citep{luhman20} astrometric missions but with larger uncertainties as expected for ground-based proper motions.

We list the proper motions for our candidate members in the last two columns of
Table \ref{tab_USco_VISTA_deepY:new_candYJ_best} in Appendix \ref{USco_VISTA_deepY:Appendix_Tables} and show the vector point diagram zoomed
on a central part in Fig.\ \ref{fig_USco_VISTA_deepY:VPD_diagram}.
The distribution of the best candidates, i.e.\ those remaining after applying filters based on $Z-J$ and $Z-K$ colours, appears to shift towards the expected mean cluster motion in the vector point diagram when compared with candidates from $YJ$ colours alone.
We observe that most of our photometric candidates lie within a 3$\sigma$ circle
centered on the mean proper motion of previous members (yellow circle in
Fig.\ \ref{fig_USco_VISTA_deepY:VPD_diagram}). Among the sources brighter than $Y$\,=\,19 mag,
we would reject 5 and 1 source(s) applying a 3$\sigma$ and 5$\sigma$ selection, respectively.
The astrometric selection would thus suggest a level of contamination of approximately 3.4--17.2\%
among the 29 sources brighter than $Y$\,=\,19 mag.
Hence, we can clear discard VISTA\,J16105178$-$233327.4 but leave the other four sources as
potential candidates. In the $Y$\,=\,19--20 mag range where the median absolute deviations
in proper motions remain below 10 mas/yr, we would reject four sources outside
3$\sigma$, one of them being outside the 5$\sigma$ limit, VISTA\,J16153235$-$2259510\@.
At faint magnitudes, we find six objects with very large proper motions but we do not
discard them as member candidates at this stage because they lie at the limit
of our deep $Y$ survey with magnitudes fainter than 21.7 mag. The median absolute
deviations increase drastically beyond $Y$\,=\,22 mag, as reported above.

%
%
\section{Spectroscopic observations}
\label{USco_VISTA_deepY:spectro}
\subsection{VLT/X-shooter near-infrared spectroscopy}
\label{USco_VISTA_deepY:spectro_XSH}
\subsubsection{Spectroscopic observations}
\label{USco_VISTA_deepY:spectro_XSH_obs}

We conducted spectroscopy of two candidates with the X-shooter \citep{dOdorico06,vernet11} 
instrument on the ESO Very Large Telescope (VLT) Unit 2 in visitor mode on the night 
of 7 May 2018 (programme 0101.C-0565; PI Lodieu).
The combination of the faintest of the targets, time allocation, and weather conditions
allowed us to collect spectra for only two of our photometric candidates with sufficient
signal-to-noise for spectral classification (Table \ref{tab_USco_VISTA_deepY:spectro_XSH_logs}).
X-shooter is a multi-wavelength cross--dispersed echelle spectrograph equipped with
three independent arms observing simultaneously in the ultraviolet (UVB;
0.3--0.56 $\mu$m), visible (VIS; 0.56--1.02 $\mu$m), and near--infrared
(NIR; 1.02--2.48 $\mu$m) whose light is split by two
dichroics. The UVB, VIS, and NIR arms are equipped with a 
4096$\times$2048 E2V CCD44-82, a 4096$\times$2048 MIT/LL
CCID\,20 detectors, and a 2096$\times$2096 Hawaii 2RG array, respectively.

We set the instrument configuration to a read-out mode of 400k with low gain 
and no binning. We employed the 1.3 arcsec slit in the UVB and 1.2 arcsec slits 
in the VIS and NIR, resulting in nominal resolving powers of 4000 (8.1 pixels per FWHM), 
6700 (7.9 pixels per FWHM), and 3900 (5.8 pixels per
FWHM) in the UVB, VIS, and NIR arms, respectively.
We set the individual integration times to 300 sec in the NIR arm and repeated 
the AB patterns multiple times to optimize the sky subtraction in the infrared. 
We carried out all observations with the slit aligned with the parallactic angle. 
We acquired both targets by pointing to a bright star located at less than 30 arcsec and applied an offset 
in both right ascension and declination calculated from the coordinates measured on the 
VISTA images. We give the logs of spectroscopic observations in 
Table \ref{tab_USco_VISTA_deepY:spectro_XSH_logs}, where we list coordinates,
the new $Y$ magnitudes, date of the observations, number of exposures, and exposure times 
set in the NIR arm for both targets.

\subsubsection{Spectroscopic data reduction}
\label{USco_VISTA_deepY:spectro_XSH_DR}

\begin{figure}
  \centering
  \includegraphics[width=\linewidth, angle=0]{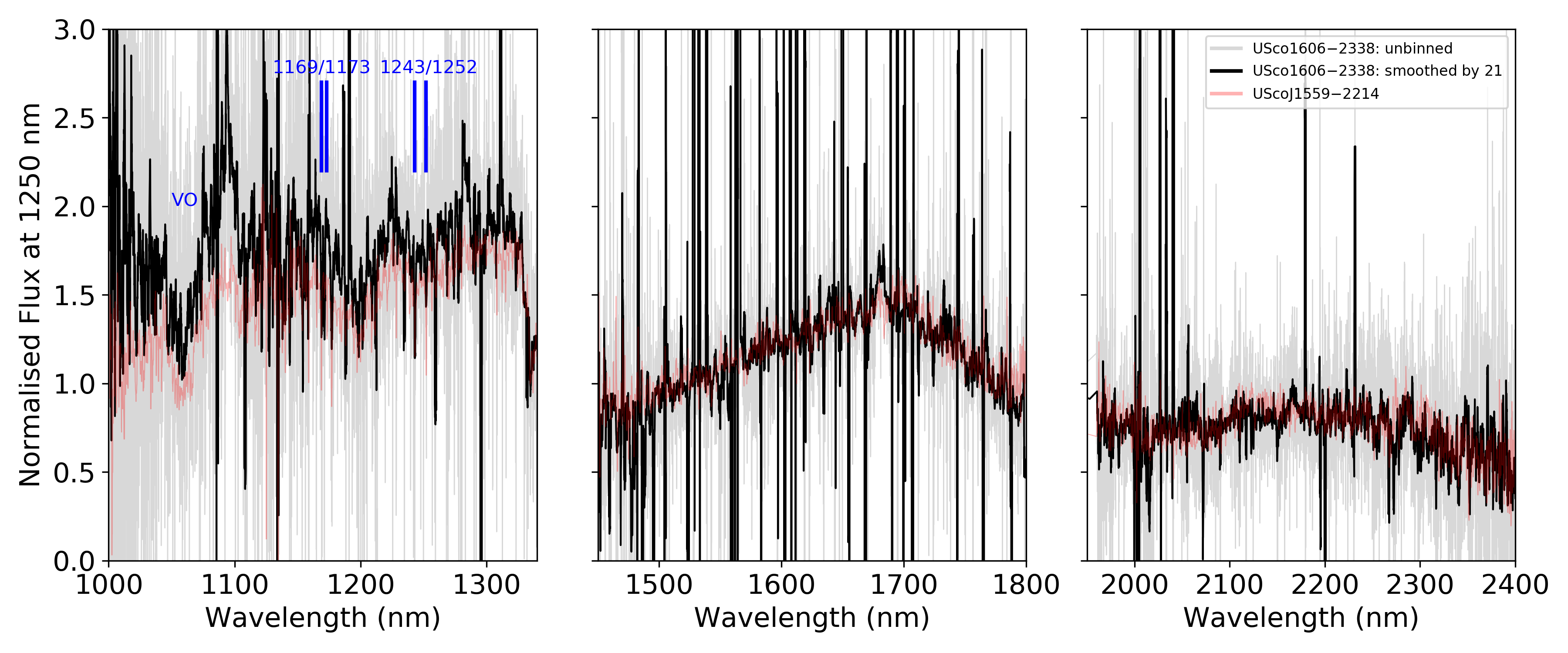}
  \includegraphics[width=\linewidth, angle=0]{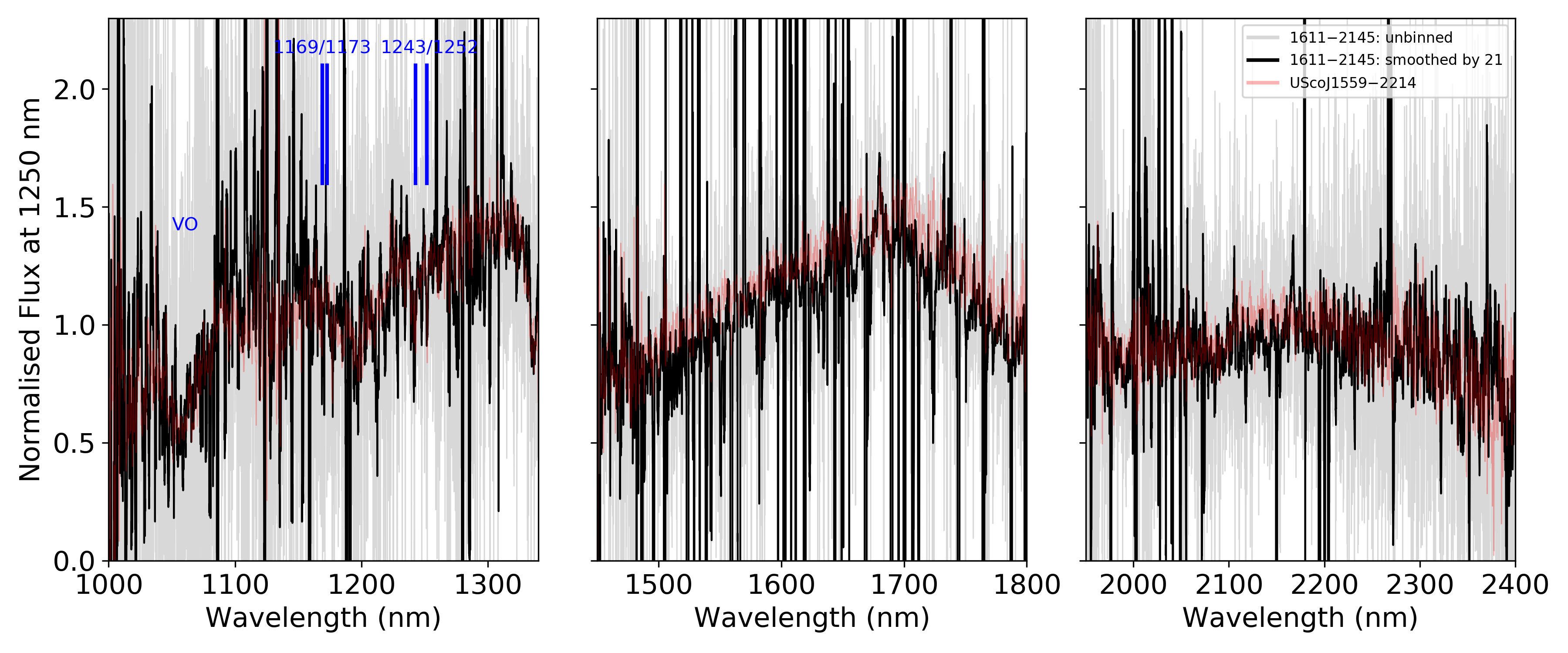}
  \caption{VLT X-shooter spectra of two of our best photometric candidates.
  Top figure: we plot USco\,16060923$-$2338114 (ID15) compared to the known 
  L1 dwarf USco\,J155936$-$221415\@.
  Bottom figure: we overplot the known L6 dwarf USco\,160413$-$224103 on top 
  of USco\,16111031$-$2145165 (ID36).
  We highlight the positions of the VO band at 1.06 microns and the two potassium doublets in the $J$-band panel, both sensitive to gravity.
}
  \label{fig_USco_VISTA_deepY:spectra_XSH}
\end{figure}

We downloaded the reduced 2D spectra of the two targets from the ESO science archive
(Table \ref{tab_USco_VISTA_deepY:spectro_XSH_logs}).
The data reduction is made automatically with the {\tt{esoreflex}} pipeline and 
includes 2D and 1D spectra extracted optimally without telluric correction.
We did not detect any signal in the UVB arm and no obvious emission line 
like H$\alpha$ in the VIS spectra of both targets, except for extremely weak signals
beyond 900 nm, hence, we concentrate our analysis on the NIR arm.
We removed tellurics from the NIR spectra of both targets with the
{\tt{molecfit}} package distributed by ESO
\citep{kausch15,smette15}\footnote{http://www.eso.org/sci/software/pipelines/skytools/molecfit}.
The final spectra are shown in Fig.\ \ref{fig_USco_VISTA_deepY:spectra_XSH}
with the smoothed spectra by a factor of 21 pixels (black) on top of the unsmoothed spectra displayed in grey.

%
%
\begin{table}
 \centering
 \caption[]{Logs of the VLT X-shooter spectroscopic observations.
We provide the identified number (Table \ref{tab_USco_VISTA_deepY:new_candYJ_best}), 
coordinates (J2000) of the targets, their $Y$-band magnitudes,
the date of observations, and the number and length of the on-source integrations 
employed for the NIR arm.
 }
 \begin{tabular}{@{\hspace{0mm}}r @{\hspace{2mm}}c @{\hspace{2mm}}c @{\hspace{2mm}}c @{\hspace{2mm}}c @{\hspace{2mm}}c @{\hspace{0mm}}}
 \hline
 \hline
ID & RA (J2000)  & dec (J2000)  &   $Y$  &     Date     & ExpT  \cr
 \hline
   & hh:mm:ss.ss &  dd:':''     &   mag  &  yyyy-mm-dd  & sec   \cr
 \hline
15 & 16:06:09.23  & $-$23:38:11.4 & 18.034$\pm$0.005 & 2018-05-07 & 6$\times$300 \cr
36 & 16:11:10.31  & $-$21:45:16.5 & 19.745$\pm$0.015 & 2018-05-07 & 12$\times$300 \cr
 \hline
 \label{tab_USco_VISTA_deepY:spectro_XSH_logs}
 \end{tabular}
\end{table}
%

%
%
\subsection{Near-infrared spectral classification}
\label{USco_VISTA_deepY:XSH_SpT}

We inspected the X-shooter infrared spectra displayed in Fig.\ \ref{fig_USco_VISTA_deepY:spectra_XSH}, which
reveal the presence of spectral features typical of young (i.e.\ low gravity) L dwarfs: 
a strong VO band at 1.06 micron, the triangular shape in the $H$-band, weak alkali lines 
like the K{\small{I}} potassium doublets at 1.169/1.177 and 1.243/1.252 microns, and strong
water bands \citep{martin96,luhman98,zapatero00,lucas00,gorlova03,mcgovern04,kirkpatrick05,cruz09,allers13,alves13a,bonnefoy14a,manjavacas14,muzic14,lodieu18a,chinchilla20a}.
We performed the spectral classification of the two sources by direct comparison with known
L0--L7 dwarf members of the USco association collected with the same VLT/X-shooter set-up
\citep{lodieu18a} and Gemini Near-Infrared Spectrograph GNIRS \citep{elias06a,lodieu08a}.
We infer infrared spectral types of L1.0$\pm$0.5 and L6$\pm$1.0 
for USco\,J16060923$-$2338114 and USco\,J16111031$-$2145165, respectively.

We investigated the strengths of the K{\small{I}} potassium doublets
present in the $J$-band region of the spectra for both objects
(left panels in Fig.\ \ref{fig_USco_VISTA_deepY:spectra_XSH}). We set upper limits
of 0.01 nm on the doublet at (1169.0,1173.3) nm. We derived pseudo-equivalent widths
of (0.031$\pm$0.005,0.019$\pm$0.04) and (0.014$\pm$0.001,0.024$\pm$0.03) nm and 
for the potassium doublet further to the red with central wavelengths of
(1242.99,1252.31) and (1243.14,1252.15) nm, for ID98600 and ID341056, respectively.
These values are in agreement with weak potassium features seen in USco L0--L7
dwarfs \citep{penya16a,lodieu18a}.

%
%
\section{Discussion and implications}
\label{USco_VISTA_deepY:discuss}
\subsection{Did we really find T-Type members?}
\label{USco_VISTA_deepY:discuss_Ts}

Old field T dwarfs are brown dwarfs whose infrared spectra are shaped by strong methane and water bands as well as H$_{2}$ collision-induced absorption
\citep{leggett00,burgasser02,geballe02,burgasser06a}. 
Their optical spectra are dominated by pressure-broadened alkali lines \citep{burgasser03d}. 
The dust present in L dwarfs has settled at the bottom of their atmospheres, resulting
in red optical and optical-to-infrared colours but blue near-infrared colours.
Their effective temperatures are typically below 1300\,K\@.

In view of the numbers of possible candidate members left after applying the difference colours cuts,
we discuss the chances that these might be T-type dwarfs belonging to USco. Based on photometry
and colours, field L and T dwarfs display constant $Y-J$ colours with some significant
scatter \citep{hewett06,lodieu07b,pinfield08,burningham10b,burningham13}, opposite to 
the sequence of USco member candidates identified in the ($Y-J$,$Y$) colour-magnitude 
diagram (Fig.\ \ref{fig_USco_VISTA_deepY:CMD_YJY}). We observe that the $Y-J$ colours
of potential late-L/early-T member candidates is not constant but rather becoming redder 
with fainter magnitudes.
This is expected for members of a cluster or association that follow a sequence, with
the coolest members exhibiting the latest spectral types.

The $Z-J$ colours of field L/T dwarfs remain relatively constant, consistent with most of our 
USco candidates. Nonetheless, we identified a few sources with distinct $Z-J$ colours compared to 
other USco candidates (Fig.\ \ref{fig_USco_VISTA_deepY:CMD_ZJZ}). On the one hand, USco\,J1611015$-$21451693 (\#17) is very red in $Z-J$ but rather blue in $Y-J$. 
On the other hand, USco\,J16051784$-$23355193 (\#6) shows red colours on both axes and
seems to extend a sequence marked by a small group of sources (\#1, 2, 7, 10, 11). 
Based on their magnitudes, \#17 is as bright as the mid-L dwarfs confirmed spectroscopically
while \#6 is among the faintest of our candidates and lies at the bottom of the sequence. 
Consequently, we suggest that one of these two new candidates might be a true T-type member in USco.
However we have neither additional photometry nor spectroscopy for it so we cannot 
yet make a strong case.

Furthermore, the sequence of old L and T dwarfs turns to the blue across the L/T transition in 
the $J-K$ colour (purple dashed line in Fig.\ \ref{fig_USco_VISTA_deepY:CMD_ZKZ_JKJ}) based on 
the absolute magnitude vs spectral type relations of \citet{dupuy12} in the Mauna Kea Observatory 
system as well as the $J-w1$ and $J-w2$ colours whereas it keeps redder in the $K-w2$ and $w1-w2$ colours (purple line in Fig.\ \ref{fig_USco_VISTA_deepY:CMD_WISE}). Unfortunately we do not observe
a kink in the USco sequence in those diagrams, most likely due to the limited depth of the
UKIDSS Galactic Clusters Survey in the $K$-band (100\% completeness of about 18.2 mag) and
the AllWISE/unWISE catalogues that do not provide any detection for our faintest candidates.
This suggests that our survey has not yet uncovered T-type planetary-mass objects in USco.

Moreover, we note that the $J$-band magnitudes of the faintest candidates are similar
because the $Y$ magnitudes become fainter and the $Y-J$ redder at the same time.
Consequently, their absolute magnitudes should be comparable, which we would expect
if the trend is consistent with field L6--T4 dwarfs having $J$-band absolute magnitudes
that differ by less than a few tens of magnitudes \citep{dupuy12}. Hence, the new
candidates might well be late-L dwarfs but could also have later spectral types (e.g.\ cases
of \#6 and \#17).

Did we really find young T-type planetary-mass members in USco? The answer is undecisive
(i.e.\ {\it{maybe}}) because we lack unambiguous evidence in spite of being two magnitudes fainter 
in $Y$ and having identified candidates one
magnitude fainter in $J$ than late-L dwarfs confirmed spectroscopically. Nonetheless, we
highlight USco\,J16051784$-$23355193 (\#6) as a potential USco T-type member due to its 
extreme $Y-J$ and $Z-J$ colours that differ from other candidates and that an increase redness.
Moreover, we cannot discard USco\,J1611015$-$21451693 (\#17) as a T-type member because it
shows red $Z-J$ colours consistent with the synthetic colours of old field T dwarfs \citep{hewett06}.
However, further photometry and spectroscopy are needed to confirm both as young T dwarfs.

\subsection{The USco IMF in the planetary-mass regime}
\label{USco_VISTA_deepY:discuss_IMF}

\begin{figure}
  \centering
  \includegraphics[width=\linewidth, angle=0]{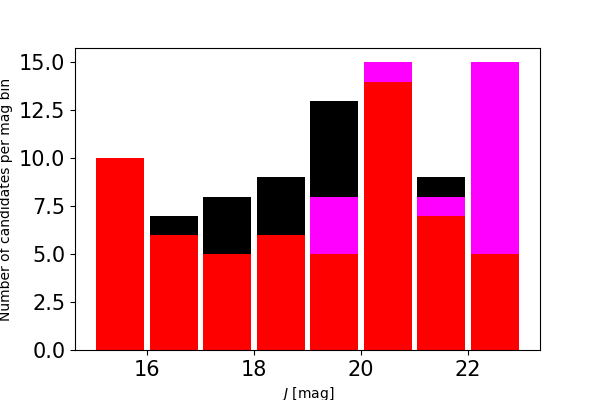}
  \caption{Luminosity function: numbers of new $YJ$ candidates in the $J$\,=\,14$-$21 mag
  interval before rejection of potential photometric non members (black). The red and magenta histograms indicate the minimum and maximum numbers of $YJ$ candidates after rejection of potential photometric non members (Table \ref{tab_USco_VISTA_deepY:numbers_LF}).
  The first and last two bins are incomplete due to saturation at bright magnitude and the 100\% completeness at $J$\,=\,20.5 mag, respectively.
  }
  \label{fig_USco_VISTA_deepY:function_dL}
\end{figure}

The aim of this subsection is to discuss the shape and form of the USco luminosity and 
mass functions into the planetary-mass regime. In Table \ref{tab_USco_VISTA_deepY:numbers_LF},
we give the final numbers of $YJ$ candidates
before and after rejection of potential photometric non members, equivalent to the 
luminosity function (Fig.\ \ref{fig_USco_VISTA_deepY:function_dL}). 
We transform the luminosity function into a mass spectrum, assuming an age of 10 Myr 
\citep{pecaut12} and the magnitude-mass relation from the BT-Settl model \citep{baraffe15}.
We also specify the mass range according to the BT-Settl 10 Myr-old isochrone in
Table \ref{tab_USco_VISTA_deepY:numbers_LF}.
We assumed an age of 10 Myr for USco based on the comparison
of these models with the masses derived from eclipsing binaries identified in the
association thanks to the Kepler K2 mission \citep{borucki10,lissauer14,batalha14}
over the past years \citep{alonso15a,kraus15a,lodieu15c,david16a,david19a}.
The BT-Settl models at 10 Myr reproduce relatively well the sequence of USco 
eclipsing binaries in the mass-radius diagram \citep{lodieu20a}.
The age of USco might well be as young as 5 Myr from isochrone fitting of the lowest
mass stars \citep{preibisch99,preibisch01,song12,david19a} with an upper limit of
about 10 Myr \citep{pecaut12}.

In Table \ref{tab_USco_VISTA_deepY:numbers_LF} we observe that the number of objects
per bin of magnitudes in the six square degrees is relatively constant over the 
$J$\,=\,14--20 mag range. We considered Poissonian errors on the luminosity
function taking the square root of the number of objects in each column
that we need to sum up to obtain the total number of candidates per magnitude bin.
This is as a very approximate but recommended practice to estimate the error bars in histograms
\footnote{\url{https://docs.astropy.org/en/stable/api/astropy.stats.poisson_conf_interval.html}}.
The last two bins are incomplete because our $J$-band survey is 100\% complete down to
$J$\,=\,20.5 mag. Typically we find 1.0--1.5 planetary-mass member per square
degree in this area of USco, up to two maximum. We observe an increase in the
planetary-mass regime if all photometric $YJ$ candidates are bona-fide members.
If only the smallest numbers of candidates is confirmed with further follow-ups,
the shape of the mass spectrum would rather be flat in the 20--7 M$_{\rm Jup}$ range.
We can discard a significant decrease in the numbers of members down to
5 M$_{\rm Jup}$, assuming an age of 10 Myr, unless all photometric candidates
are rejected in the future. If the age of USco is 5 Myr, 
the lower mass limit of our survey would be just below 4 M$_{\rm Jup}$.
In spite of all uncertainties associated to models at these low masses and
young ages as well as the lack of spectroscopic confirmation, star formation
processes can form isolated planetary-mass objects down to 5 M$_{\rm Jup}$,
consistent with the findings in the solar neighbourhood \citep{kirkpatrick19}.
%
%
\begin{table*}
 \centering
 \caption[]{
 Numbers of new $YJ$ candidates per bin of $J$ magnitudes.
 We list the total number of objects (=\,luminosity function)
 in the area of six square degrees covered by our deep $Y$ survey.
 We give the numbers of $YJ$ candidates before ($YJ$) and after rejection (Rej)
 based on their position in diagrams involving additional passbands.
 We also list the four candidates undetected in the VISTA $Z$ survey (no$Z$)
 and those remaining as potential candidates from the deep HSC survey (HSC$Z$).
 We take into account that the Subaru survey covers only 20\% of 
 the VISTA survey, resulting in a factor of five to consider when counting the
 number of potential candidates after applying the selection in the $z$-band.
 The final number (or range) of objects (dN) is the sum of columns 2, 4, and 5
 minus the number of rejected candidates in column 3\@. We give in parenthesis
 the range of values considering the Poissonian statistics rounded to the
 nearest decimal.
 The mass range is computed using the BT-Settl 10 Myr-old isochrones.
 The last column gives the number of objects per mass bin rounded to the nearest
 integer, dividing dN by the difference of the mass range with its associated Poissonian 
 interval in parenthesis.
 We warn that the first and last two bins are incomplete due to saturation
 at bright magnitude and the 100\% completeness at $J$\,=\,20.5 mag,
 respectively. 
 }
 \begin{tabular}{@{\hspace{0mm}}c @{\hspace{2mm}}c @{\hspace{2mm}}c @{\hspace{2mm}}c @{\hspace{2mm}}c@{\hspace{2mm}}c @{\hspace{2mm}}c @{\hspace{2mm}}c @{\hspace{0mm}}}
 \hline
 \hline
$J$     &  $YJ$ &   Rej  &   no$Z$   & HSC\,$Z$     &   dN    & Mass range  & dN/dM \cr
 \hline
mag     &       &        &           &            &          &  M$_{\odot}$ &  \cr
 \hline
14$-$15 &  10   &     0  &     0     &    0       & 10  (6.8--13.1)   & 0.0600--0.0220  &  263 (247--279) \cr
15$-$16 &   7   &     1  &     0     &    0       &  6  (3.5--8.5)    & 0.0220--0.0160  & 1000 (968--1032) \cr
16$-$17 &   8   &     3  &     0     &    0       &  5  (2.7--7.2)    & 0.0160--0.0120  & 1250 (1214--1285) \cr
17$-$18 &   9   &     3  &     0     &    0       &  6   (3.5--8.5)   & 0.0120--0.0095  & 2400 (2351--2449) \cr
18$-$19 &  13   &  5--8  &     0     &    0       & 5--8 (2.7--10.3)  & 0.0095--0.0073  & 2242--3587 (2194--3647) \cr
19$-$20 &   9   &     0  &  0--1     & 1$\times$5 & 14--15 (8.1--20.9) & 0.0073--0.0059  & 10000--10714 (5786-14929) \cr
20$-$21 &   1   &     0  &  1--2     & 1$\times$5 &  7--8 (1.5--13.5)  & 0.0059--0.0045 & 5000--5714 (1071-9643) \cr
21$-$22 &   0   &     0  &     0     & (2$\pm$1)$\times$5 & $\geq$(5--15) $\geq$(2.9--17.1) & 0.0045--0.0037 & $\geq$6250--18750 (3625-21375) \cr
 \hline
 \label{tab_USco_VISTA_deepY:numbers_LF}
 \end{tabular}
\end{table*}
%

%
%
%

%
%
\section{Conclusions and future work}

We presented a dedicated photometric search in $\sim$6 deg$^{2}$ in the central region 
of the USco association to look for the coolest members with a special focus on late-L and
T-type candidates. Our survey is the deepest in the association over such large area,
between 2 and 1 magnitude deeper in $Y$ and $J$ than any previous study and we are now
closer to reach the regime of T dwarfs in USco.
The main results of our study are: 

\begin{itemize}
\item [$\bullet$] We identified 10 new candidates fainter than the previously known L7
members confirmed spectroscopically. Their magnitudes are fainter than $Y$\,=\,21.7 mag ($J$\,$>$\,19.5 mag),
equivalent to masses below 5 M$_{\rm Jup}$ according to evolutionary models at an age of 10 Myr.
\item [$\bullet$] We highlight two potential T-type member candidates among the 10 faintest sources identified in our deep survey.
\item [$\bullet$] We derived proper motion confirmation of the brightest member candidates
while the accuracy of the astrometry is limited for the newest and faintest candidates.
\item [$\bullet$] We presented near-infrared spectroscopy of two photometric candidates
confirmed them as young L dwarfs, adding credence to their membership. The remaining candidates
require optical and/or infrared spectroscopic follow-up.
\item [$\bullet$] We derive photometric estimates of the luminosity function 
and mass spectrum in the planetary-mass domain showing no obvious sign of dearth
of members down to masses of 5 M$_{\rm Jup}$. 
Although this analysis is preliminary and requires photometric and spectroscopic
follow-ups, there is now growing observational evidence that the star formation processes
can form isolated objects with masses below 5 M$_{\rm Jup}$ in the field and in young
regions. 
\end{itemize}

%
%
\section*{Acknowledgments}
NL was partly funded by the programme AYA2015-69350-C3-2-P from Spanish Ministry of Economy and Competitiveness (MINECO) and acknowledges  support  from  the Agencia  Estatal  de  Investigaci\'on  del  Ministerio  de  Ciencia  e Innovaci\'on (AEI-MCINN) under grant PID2019-109522GB-C53\@.
NL has benefited from internal funding from an IAC Severo Ochoa outgoing fellowship for a stay at the Royal Observatory in Edinburgh.
We warmly thank the staff at the Royal Observatory Edinburgh and Cambridge Astronomy Survey Unit 
for their valuable help and support with the data reduction, in particular Mike Read, Rob Blake,
Eckhard Sutorius, Mike Irwin, Aybuke Kupcu Yoldas, and Carlos Gonz\'alez-Ferna\'ndez.

This research has made use of the Simbad and Vizier \citep{ochsenbein00}
databases, operated at the Centre de Donn\'ees Astronomiques de Strasbourg
(CDS), and of NASA's Astrophysics Data System Bibliographic Services (ADS).

Based on observations collected at the European Organisation for Astronomical Research 
in the Southern Hemisphere under ESO programme(s) 095.C-0781(A), 089-C.0102(ABC), and 0101.C-0565\@.

This work is based on programmes GTC4-14A (PI Lodieu)
made with the Gran Telescopio Canarias (GTC), operated on the island of La Palma
in the Spanish Observatorio del Roque de los Muchachos of the Instituto de Astrof\'isica de Canarias.

This work is partly based on data collected at Subaru Telescope, which is operated by the National Astronomical Observatory of Japan.

Based on data from the UKIRT Infrared Deep Sky Survey (UKIDSS). The UKIDSS project 
is defined in \citet{lawrence07} and uses the UKIRT Wide Field Camera \citep[WFCAM;][]{casali07}. 
The photometric system is described in \citet{hewett06} and the calibration is described 
\citet{hodgkin09}. The pipeline processing and science archive are described in 
Irwin et al. (2009, in prep) and \citet{hambly08}.

The Pan-STARRS1 Surveys (PS1) have been made possible through contributions of the Institute for Astronomy, the University of Hawaii, the Pan-STARRS Project Office, the Max-Planck Society and its participating institutes, the Max Planck Institute for Astronomy, Heidelberg and the Max Planck Institute for Extraterrestrial Physics, Garching, The Johns Hopkins University, Durham University, the University of Edinburgh, Queen's University Belfast, the Harvard-Smithsonian Center for Astrophysics, the Las Cumbres Observatory Global Telescope Network Incorporated, the National Central University of Taiwan, the Space Telescope Science Institute, the National Aeronautics and Space Administration under Grant No. NNX08AR22G issued through the Planetary Science Division of the NASA Science Mission Directorate, the National Science Foundation under Grant No. AST-1238877, the University of Maryland, and Eotvos Lorand University (ELTE).

This publication makes use of data products from the Wide-field Infrared Survey Explorer, which 
is a joint project of the University of California, Los Angeles, and the Jet Propulsion Laboratory/California 
Institute of Technology, and NEOWISE, which is a project of the Jet Propulsion Laboratory/California 
Institute of Technology. WISE and NEOWISE are funded by the National Aeronautics and Space Administration.

John D.\ Hunter.\ Matplotlib: A 2D Graphics Environment, Computing in Science \& Engineering, 9, 90-95 (2007), DOI:10.1109/MCSE.2007.55 (publisher link)
Stefan van der Walt, S.\ Chris Colbert and Ga\"el Varoquaux. The NumPy Array: A Structure for Efficient Numerical Computation, Computing in Science \& Engineering, 13, 22-30 (2011), DOI:10.1109/MCSE.2011.37 (publisher link)
Jones E, Oliphant E, Peterson P, et al.\ SciPy: Open Source Scientific Tools for Python, 2001-, http://www.scipy.org/ [Online; accessed 2017-10-03].

TOPCAT was initially (2003-2005) developed under the UK Starlink project (1980-2005, R.I.P.). 
Since then it has been supported by grant PP/D002486/1 from the UK's Particle Physics and 
Astronomy Research Council, the VOTech project (from EU FP6), the AstroGrid project (from 
PPARC/STFC), the AIDA project (from EU FP7), grants ST/H008470/1, ST/I00176X/1, ST/J001414/1 
and ST/L002388/1 from the UK's Science and Technology Facilities Council (STFC), the GAVO 
project (BMBF Bewilligungsnummer 05A08VHA), the European Space Agency, and the FP7 
project GENIUS\@. 

%
%
\section*{Data Availability}
The data underlying this article will be made public to the community through Vizier at the Centre de Donn\'ees de Strasbourg at \url{http://cdsweb.u-strasbg.fr}, and can be accessed with the bibcode or last name of the first author of the paper.

%
%
\bibliographystyle{mnras}
\bibliography{mnemonic,biblio_old,nch}

\begin{thebibliography}{}
\makeatletter
\relax
\def\mn@urlcharsother{\let\do\@makeother \do\$\do\&\do\#\do\^\do\_\do\%\do\~}
\def\mn@doi{\begingroup\mn@urlcharsother \@ifnextchar [ {\mn@doi@}
  {\mn@doi@[]}}
\def\mn@doi@[#1]#2{\def\@tempa{#1}\ifx\@tempa\@empty \href
  {http://dx.doi.org/#2} {doi:#2}\else \href {http://dx.doi.org/#2} {#1}\fi
  \endgroup}
\def\mn@eprint#1#2{\mn@eprint@#1:#2::\@nil}
\def\mn@eprint@arXiv#1{\href {http://arxiv.org/abs/#1} {{\tt arXiv:#1}}}
\def\mn@eprint@dblp#1{\href {http://dblp.uni-trier.de/rec/bibtex/#1.xml}
  {dblp:#1}}
\def\mn@eprint@#1:#2:#3:#4\@nil{\def\@tempa {#1}\def\@tempb {#2}\def\@tempc
  {#3}\ifx \@tempc \@empty \let \@tempc \@tempb \let \@tempb \@tempa \fi \ifx
  \@tempb \@empty \def\@tempb {arXiv}\fi \@ifundefined
  {mn@eprint@\@tempb}{\@tempb:\@tempc}{\expandafter \expandafter \csname
  mn@eprint@\@tempb\endcsname \expandafter{\@tempc}}}

\bibitem[\protect\citeauthoryear{{Allers} \& {Liu}}{{Allers} \&
  {Liu}}{2013}]{allers13}
{Allers} K.~N.,  {Liu} M.~C.,  2013, \mn@doi [ApJ]
  {10.1088/0004-637X/772/2/79}, \href
  {http://cdsads.u-strasbg.fr/abs/2013ApJ...772...79A} {772, 79}

\bibitem[\protect\citeauthoryear{{Alonso}, {Deeg}, {Hoyer}, {Lodieu}, {Palle}
  \& {Sanchis-Ojeda}}{{Alonso} et~al.}{2015}]{alonso15a}
{Alonso} R.,  {Deeg} H.~J.,  {Hoyer} S.,  {Lodieu} N.,  {Palle} E.,
  {Sanchis-Ojeda} R.,  2015, \mn@doi [A\&A] {10.1051/0004-6361/201527109},
  \href {http://cdsads.u-strasbg.fr/abs/2015A%26A...584L...8A} {584, L8}

\bibitem[\protect\citeauthoryear{{Alves de Oliveira}, {Moraux}, {Bouvier},
  {Duch{\^e}ne}, {Bouy}, {Maschberger}  \& {Hudelot}}{{Alves de Oliveira}
  et~al.}{2013}]{alves13a}
{Alves de Oliveira} C.,  {Moraux} E.,  {Bouvier} J.,  {Duch{\^e}ne} G.,  {Bouy}
  H.,  {Maschberger} T.,   {Hudelot} P.,  2013, \mn@doi [A\&A]
  {10.1051/0004-6361/201220229}, \href
  {http://cdsads.u-strasbg.fr/abs/2013A%26A...549A.123A} {549, A123}

\bibitem[\protect\citeauthoryear{{Ardila}, {Mart{\'{\i}}n}  \&
  {Basri}}{{Ardila} et~al.}{2000}]{ardila00}
{Ardila} D.,  {Mart{\'{\i}}n} E.,   {Basri} G.,  2000, AJ, \href
  {http://cdsads.u-strasbg.fr/cgi-bin/nph-bib_query?bibcode=2000AJ....120..479A&db_key=AST}
  {120, 479}

\bibitem[\protect\citeauthoryear{{Baraffe}, {Homeier}, {Allard}  \&
  {Chabrier}}{{Baraffe} et~al.}{2015}]{baraffe15}
{Baraffe} I.,  {Homeier} D.,  {Allard} F.,   {Chabrier} G.,  2015, \mn@doi
  [A\&A] {10.1051/0004-6361/201425481}, \href
  {http://cdsads.u-strasbg.fr/abs/2015A%26A...577A..42B} {577, A42}

\bibitem[\protect\citeauthoryear{{Barsony}, {Haisch}, {Marsh}  \&
  {McCarthy}}{{Barsony} et~al.}{2012}]{barsony12}
{Barsony} M.,  {Haisch} K.~E.,  {Marsh} K.~A.,   {McCarthy} C.,  2012, \mn@doi
  [ApJ] {10.1088/0004-637X/751/1/22}, \href
  {http://cdsads.u-strasbg.fr/abs/2012ApJ...751...22B} {751, 22}

\bibitem[\protect\citeauthoryear{{Bastian}, {Covey}  \& {Meyer}}{{Bastian}
  et~al.}{2010}]{bastian10}
{Bastian} N.,  {Covey} K.~R.,   {Meyer} M.~R.,  2010, \mn@doi [ARA\&A]
  {10.1146/annurev-astro-082708-101642}, \href
  {http://cdsads.u-strasbg.fr/abs/2010ARA%26A..48..339B} {48, 339}

\bibitem[\protect\citeauthoryear{{Batalha}}{{Batalha}}{2014}]{batalha14}
{Batalha} N.~M.,  2014, \mn@doi [Proceedings of the National Academy of
  Science] {10.1073/pnas.1304196111}, \href
  {http://adsabs.harvard.edu/abs/2014PNAS..11112647B} {111, 12647}

\bibitem[\protect\citeauthoryear{{B{\'e}jar}, {Zapatero Osorio},
  {P{\'e}rez-Garrido}, {{\'A}lvarez}, {Mart{\'{\i}}n}, {Rebolo},
  {Vill{\'o}-P{\'e}rez}  \& {D{\'{\i}}az-S{\'a}nchez}}{{B{\'e}jar}
  et~al.}{2008}]{bejar08}
{B{\'e}jar} V.~J.~S.,  {Zapatero Osorio} M.~R.,  {P{\'e}rez-Garrido} A.,
  {{\'A}lvarez} C.,  {Mart{\'{\i}}n} E.~L.,  {Rebolo} R.,
  {Vill{\'o}-P{\'e}rez} I.,   {D{\'{\i}}az-S{\'a}nchez} A.,  2008, \mn@doi
  [ApJL] {10.1086/527557}, \href
  {http://cdsads.u-strasbg.fr/abs/2008ApJ...673L.185B} {673, L185}

\bibitem[\protect\citeauthoryear{{Best} et~al.,}{{Best} et~al.}{2017}]{best17a}
{Best} W.~M.~J.,  et~al., 2017, \mn@doi [ApJ] {10.3847/1538-4357/aa5df0}, \href
  {http://cdsads.u-strasbg.fr/abs/2017ApJ...837...95B} {837, 95}

\bibitem[\protect\citeauthoryear{{Boley} \& {Durisen}}{{Boley} \&
  {Durisen}}{2010}]{boley10}
{Boley} A.~C.,  {Durisen} R.~H.,  2010, \mn@doi [ApJ]
  {10.1088/0004-637X/724/1/618}, \href
  {http://cdsads.u-strasbg.fr/abs/2010ApJ...724..618B} {724, 618}

\bibitem[\protect\citeauthoryear{{Bonnefoy}, {Chauvin}, {Lagrange}, {Rojo},
  {Allard}, {Pinte}, {Dumas}  \& {Homeier}}{{Bonnefoy}
  et~al.}{2014}]{bonnefoy14a}
{Bonnefoy} M.,  {Chauvin} G.,  {Lagrange} A.-M.,  {Rojo} P.,  {Allard} F.,
  {Pinte} C.,  {Dumas} C.,   {Homeier} D.,  2014, \mn@doi [A\&A]
  {10.1051/0004-6361/201118270}, \href
  {http://cdsads.u-strasbg.fr/abs/2014A%26A...562A.127B} {562, A127}

\bibitem[\protect\citeauthoryear{{Borucki} et~al.,}{{Borucki}
  et~al.}{2010}]{borucki10}
{Borucki} W.~J.,  et~al., 2010, \mn@doi [Science] {10.1126/science.1185402},
  \href {http://cdsads.u-strasbg.fr/abs/2010Sci...327..977B} {327, 977}

\bibitem[\protect\citeauthoryear{{Bosch} et~al.}{{Bosch}
  et~al.}{2018}]{bosch2018}
{Bosch} J.,  et~al., 2018, \mn@doi [PASJ] {10.1093/pasj/psx080}, \href
  {https://ui.adsabs.harvard.edu/abs/2018PASJ...70S...5B} {70, S5}

\bibitem[\protect\citeauthoryear{{Boss}}{{Boss}}{1988}]{boss88}
{Boss} A.~P.,  1988, \mn@doi [ApJ] {10.1086/166563}, \href
  {http://cdsads.u-strasbg.fr/abs/1988ApJ...331..370B} {331, 370}

\bibitem[\protect\citeauthoryear{{Boyd} \& {Whitworth}}{{Boyd} \&
  {Whitworth}}{2005}]{boyd05}
{Boyd} D.~F.~A.,  {Whitworth} A.~P.,  2005, \mn@doi [A\&A]
  {10.1051/0004-6361:20041703}, \href
  {http://cdsads.u-strasbg.fr/abs/2005A%26A...430.1059B} {430, 1059}

\bibitem[\protect\citeauthoryear{{Burgasser} et~al.,}{{Burgasser}
  et~al.}{2002}]{burgasser02}
{Burgasser} A.~J.,  et~al., 2002, ApJ, \href
  {http://cdsads.u-strasbg.fr/cgi-bin/nph-bib_query?bibcode=2002ApJ...564..421B&db_key=AST}
  {564, 421}

\bibitem[\protect\citeauthoryear{{Burgasser}, {Kirkpatrick}, {Liebert}  \&
  {Burrows}}{{Burgasser} et~al.}{2003}]{burgasser03d}
{Burgasser} A.~J.,  {Kirkpatrick} J.~D.,  {Liebert} J.,   {Burrows} A.,  2003,
  ApJ, \href
  {http://cdsads.u-strasbg.fr/cgi-bin/nph-bib_query?bibcode=2003ApJ...594..510B&amp;db_key=AST}
  {594, 510}

\bibitem[\protect\citeauthoryear{{Burgasser}, {Geballe}, {Leggett},
  {Kirkpatrick}  \& {Golimowski}}{{Burgasser} et~al.}{2006}]{burgasser06a}
{Burgasser} A.~J.,  {Geballe} T.~R.,  {Leggett} S.~K.,  {Kirkpatrick} J.~D.,
  {Golimowski} D.~A.,  2006, \mn@doi [ApJ] {10.1086/498563}, \href
  {http://cdsads.u-strasbg.fr/cgi-bin/nph-bib_query?bibcode=2006ApJ...637.1067B&db_key=AST}
  {637, 1067}

\bibitem[\protect\citeauthoryear{{Burgess}, {Moraux}, {Bouvier}, {Marmo},
  {Albert}  \& {Bouy}}{{Burgess} et~al.}{2009}]{burgess09}
{Burgess} A.~S.~M.,  {Moraux} E.,  {Bouvier} J.,  {Marmo} C.,  {Albert} L.,
  {Bouy} H.,  2009, \mn@doi [A\&A] {10.1051/0004-6361/200912444}, \href
  {http://cdsads.u-strasbg.fr/abs/2009A%26A...508..823B} {508, 823}

\bibitem[\protect\citeauthoryear{{Burningham} et~al.,}{{Burningham}
  et~al.}{2010}]{burningham10b}
{Burningham} B.,  et~al., 2010, \mn@doi [MNRAS]
  {10.1111/j.1365-2966.2010.16800.x}, \href
  {http://cdsads.u-strasbg.fr/abs/2010MNRAS.406.1885B} {406, 1885}

\bibitem[\protect\citeauthoryear{{Burningham} et~al.,}{{Burningham}
  et~al.}{2013}]{burningham13}
{Burningham} B.,  et~al., 2013, \mn@doi [MNRAS] {10.1093/mnras/stt740}, \href
  {http://cdsads.u-strasbg.fr/abs/2013MNRAS.433..457B} {433, 457}

\bibitem[\protect\citeauthoryear{{Casali}, {Adamson}, {Alves de Oliveira},
  {Almaini}, {Burch}, {Chuter}, {Elliot}  \& {23 co-authors}}{{Casali}
  et~al.}{2007}]{casali07}
{Casali} M.,  {Adamson} A.,  {Alves de Oliveira} C.,  {Almaini} O.,  {Burch}
  K.,  {Chuter} T.,  {Elliot} J.,   {23 co-authors} 2007, \mn@doi [A\&A]
  {10.1051/0004-6361:20066514}, \href
  {http://cdsads.u-strasbg.fr/abs/2007A%26A...467..777C} {467, 777}

\bibitem[\protect\citeauthoryear{{Cepa} et~al.,}{{Cepa} et~al.}{2000}]{cepa00}
{Cepa} J.,  et~al., 2000, in {M.~Iye \& A.~F.~Moorwood} ed.,  Society of
  Photo-Optical Instrumentation Engineers (SPIE) Conference Series Vol. 4008,
  Society of Photo-Optical Instrumentation Engineers (SPIE) Conference Series.
  pp 623--631

\bibitem[\protect\citeauthoryear{{Chambers}, {Magnier}, {Metcalfe}  \& {103
  co-authors}}{{Chambers} et~al.}{2016}]{chambers16a}
{Chambers} K.~C.,  {Magnier} E.~A.,  {Metcalfe} N.,   {103 co-authors} 2016,
  ApJ, \href {http://adsabs.harvard.edu/abs/2016arXiv161205560C} {}

\bibitem[\protect\citeauthoryear{{Chiang} \& {Chen}}{{Chiang} \&
  {Chen}}{2015}]{chiang15b}
{Chiang} P.,  {Chen} W.~P.,  2015, \mn@doi [ApJL]
  {10.1088/2041-8205/811/2/L16}, \href
  {http://cdsads.u-strasbg.fr/abs/2015ApJ...811L..16C} {811, L16}

\bibitem[\protect\citeauthoryear{{Chinchilla} et~al.,}{{Chinchilla}
  et~al.}{2020}]{chinchilla20a}
{Chinchilla} P.,  et~al., 2020, \mn@doi [A\&A] {10.1051/0004-6361/201936130},
  \href {https://ui.adsabs.harvard.edu/abs/2020A&A...633A.152C} {633, A152}

\bibitem[\protect\citeauthoryear{{Cook}, {Scholz}  \& {Jayawardhana}}{{Cook}
  et~al.}{2017}]{cook17a}
{Cook} N.~J.,  {Scholz} A.,   {Jayawardhana} R.,  2017, \mn@doi [AJ]
  {10.3847/1538-3881/aa9751}, \href
  {https://ui.adsabs.harvard.edu/abs/2017AJ....154..256C} {154, 256}

\bibitem[\protect\citeauthoryear{{Cross} et~al.,}{{Cross}
  et~al.}{2012}]{cross12}
{Cross} N.~J.~G.,  et~al., 2012, \mn@doi [A\&A] {10.1051/0004-6361/201219505},
  \href {http://cdsads.u-strasbg.fr/abs/2012A%26A...548A.119C} {548, A119}

\bibitem[\protect\citeauthoryear{{Cruz}, {Kirkpatrick}  \& {Burgasser}}{{Cruz}
  et~al.}{2009}]{cruz09}
{Cruz} K.~L.,  {Kirkpatrick} J.~D.,   {Burgasser} A.~J.,  2009, \mn@doi [AJ]
  {10.1088/0004-6256/137/2/3345}, \href
  {http://cdsads.u-strasbg.fr/abs/2009AJ....137.3345C} {137, 3345}

\bibitem[\protect\citeauthoryear{{Cutri} \& {et al.}}{{Cutri} \& {et
  al.}}{2014}]{cutri14}
{Cutri} R.~M.,  {et al.} 2014, VizieR Online Data Catalog, \href
  {http://cdsads.u-strasbg.fr/abs/2014yCat.2328....0C} {2328, 0}

\bibitem[\protect\citeauthoryear{{Cutri} et~al.,}{{Cutri}
  et~al.}{2013}]{cutri13}
{Cutri} R.~M.,  et~al., 2013, {Explanatory Supplement to the AllWISE Data
  Release Products}, Explanatory Supplement to the AllWISE Data Release
  Products

\bibitem[\protect\citeauthoryear{{D'Odorico} et~al.,}{{D'Odorico}
  et~al.}{2006}]{dOdorico06}
{D'Odorico} S.,  et~al., 2006, in Society of Photo-Optical Instrumentation
  Engineers (SPIE) Conference Series. , \mn@doi{10.1117/12.672969}

\bibitem[\protect\citeauthoryear{{Dalton} et~al.,}{{Dalton}
  et~al.}{2006}]{dalton06}
{Dalton} G.~B.,  et~al., 2006, in Society of Photo-Optical Instrumentation
  Engineers (SPIE) Conference Series. , \mn@doi{10.1117/12.670018}

\bibitem[\protect\citeauthoryear{{David} et~al.,}{{David}
  et~al.}{2016a}]{david16b}
{David} T.~J.,  et~al., 2016a, \mn@doi [Nat] {10.1038/nature18293}, \href
  {http://cdsads.u-strasbg.fr/abs/2016Natur.534..658D} {534, 658}

\bibitem[\protect\citeauthoryear{{David}, {Hillenbrand}, {Cody}, {Carpenter}
  \& {Howard}}{{David} et~al.}{2016b}]{david16a}
{David} T.~J.,  {Hillenbrand} L.~A.,  {Cody} A.~M.,  {Carpenter} J.~M.,
  {Howard} A.~W.,  2016b, \mn@doi [ApJ] {10.3847/0004-637X/816/1/21}, \href
  {http://cdsads.u-strasbg.fr/abs/2016ApJ...816...21D} {816, 21}

\bibitem[\protect\citeauthoryear{{David}, {Hillenbrand}, {Gillen}, {Cody},
  {Howell}, {Isaacson}  \& {Livingston}}{{David} et~al.}{2019}]{david19a}
{David} T.~J.,  {Hillenbrand} L.~A.,  {Gillen} E.,  {Cody} A.~M.,  {Howell}
  S.~B.,  {Isaacson} H.~T.,   {Livingston} J.~H.,  2019, \mn@doi [ApJ]
  {10.3847/1538-4357/aafe09}, \href
  {https://ui.adsabs.harvard.edu/abs/2019ApJ...872..161D} {872, 161}

\bibitem[\protect\citeauthoryear{{Davis}}{{Davis}}{1999}]{davis99a}
{Davis} L.~E.,  1999, in {Craine} E.~R.,  {Crawford} D.~L.,   {Tucker} R.~A.,
  eds,  Astronomical Society of the Pacific Conference Series Vol. 189,
  Precision CCD Photometry. p.~35

\bibitem[\protect\citeauthoryear{{Dawson}, {Scholz}  \& {Ray}}{{Dawson}
  et~al.}{2011}]{dawson11}
{Dawson} P.,  {Scholz} A.,   {Ray} T.~P.,  2011, \mn@doi [MNRAS]
  {10.1111/j.1365-2966.2011.19573.x}, \href
  {http://cdsads.u-strasbg.fr/abs/2011MNRAS.418.1231D} {418, 1231}

\bibitem[\protect\citeauthoryear{{Dawson}, {Scholz}, {Ray}, {Marsh}, {Wood},
  {Natta}, {Padgett}  \& {Ressler}}{{Dawson} et~al.}{2013}]{dawson13}
{Dawson} P.,  {Scholz} A.,  {Ray} T.~P.,  {Marsh} K.~A.,  {Wood} K.,  {Natta}
  A.,  {Padgett} D.,   {Ressler} M.~E.,  2013, \mn@doi [MNRAS]
  {10.1093/mnras/sts386}, \href
  {http://cdsads.u-strasbg.fr/abs/2013MNRAS.429..903D} {429, 903}

\bibitem[\protect\citeauthoryear{{Dawson}, {Scholz}, {Ray}, {Peterson},
  {Rodgers-Lee}  \& {Geers}}{{Dawson} et~al.}{2014}]{dawson14}
{Dawson} P.,  {Scholz} A.,  {Ray} T.~P.,  {Peterson} D.~E.,  {Rodgers-Lee} D.,
   {Geers} V.,  2014, \mn@doi [MNRAS] {10.1093/mnras/stu973}, \href
  {http://cdsads.u-strasbg.fr/abs/2014MNRAS.442.1586D} {442, 1586}

\bibitem[\protect\citeauthoryear{{Dupuy} \& {Liu}}{{Dupuy} \&
  {Liu}}{2012}]{dupuy12}
{Dupuy} T.~J.,  {Liu} M.~C.,  2012, \mn@doi [ApJS]
  {10.1088/0067-0049/201/2/19}, \href
  {http://cdsads.u-strasbg.fr/abs/2012ApJS..201...19D} {201, 19}

\bibitem[\protect\citeauthoryear{{Elias}, {Joyce}, {Liang}, {Muller}, {Hileman}
   \& {George}}{{Elias} et~al.}{2006}]{elias06a}
{Elias} J.~H.,  {Joyce} R.~R.,  {Liang} M.,  {Muller} G.~P.,  {Hileman} E.~A.,
   {George} J.~R.,  2006, in Ground-based and Airborne Instrumentation for
  Astronomy. Edited by McLean, Ian S.; Iye, Masanori. Proceedings of the SPIE,
  Volume 6269, pp. (2006).. , \mn@doi{10.1117/12.671817}

\bibitem[\protect\citeauthoryear{{Emerson}}{{Emerson}}{2001}]{emerson01}
{Emerson} J.~P.,  2001, in {R.~Clowes, A.~Adamson, \& G.~Bromage} ed.,
  Astronomical Society of the Pacific Conference Series Vol. 232, The New Era
  of Wide Field Astronomy. p.~339

\bibitem[\protect\citeauthoryear{{Emerson}, {Sutherland}, {McPherson}, {Craig},
  {Dalton}  \& {Ward}}{{Emerson} et~al.}{2004}]{emerson04}
{Emerson} J.~P.,  {Sutherland} W.~J.,  {McPherson} A.~M.,  {Craig} S.~C.,
  {Dalton} G.~B.,   {Ward} A.~K.,  2004, The Messenger, \href
  {http://cdsads.u-strasbg.fr/abs/2004Msngr.117...27E} {117, 27}

\bibitem[\protect\citeauthoryear{{Faherty} et~al.,}{{Faherty}
  et~al.}{2016}]{faherty16a}
{Faherty} J.~K.,  et~al., 2016, \mn@doi [ApJS] {10.3847/0067-0049/225/1/10},
  \href {http://cdsads.u-strasbg.fr/abs/2016ApJS..225...10F} {225, 10}

\bibitem[\protect\citeauthoryear{{Filippazzo}, {Rice}, {Faherty}, {Cruz}, {Van
  Gordon}  \& {Looper}}{{Filippazzo} et~al.}{2015}]{filippazzo15}
{Filippazzo} J.~C.,  {Rice} E.~L.,  {Faherty} J.,  {Cruz} K.~L.,  {Van Gordon}
  M.~M.,   {Looper} D.~L.,  2015, \mn@doi [ApJ] {10.1088/0004-637X/810/2/158},
  \href {http://cdsads.u-strasbg.fr/abs/2015ApJ...810..158F} {810, 158}

\bibitem[\protect\citeauthoryear{{Forgan} \& {Rice}}{{Forgan} \&
  {Rice}}{2011}]{forgan11}
{Forgan} D.,  {Rice} K.,  2011, \mn@doi [MNRAS]
  {10.1111/j.1365-2966.2011.19380.x}, \href
  {http://cdsads.u-strasbg.fr/abs/2011MNRAS.417.1928F} {417, 1928}

\bibitem[\protect\citeauthoryear{{Furusawa} et~al.,}{{Furusawa}
  et~al.}{2018}]{furusawa18a}
{Furusawa} H.,  et~al., 2018, \mn@doi [PASJ] {10.1093/pasj/psx079}, \href
  {https://ui.adsabs.harvard.edu/abs/2018PASJ...70S...3F} {70, S3}

\bibitem[\protect\citeauthoryear{{Gagn{\'e}} et~al.,}{{Gagn{\'e}}
  et~al.}{2015a}]{gagne15c}
{Gagn{\'e}} J.,  et~al., 2015a, \mn@doi [ApJS] {10.1088/0067-0049/219/2/33},
  \href {http://cdsads.u-strasbg.fr/abs/2015ApJS..219...33G} {219, 33}

\bibitem[\protect\citeauthoryear{{Gagn{\'e}}, {Lafreni{\`e}re}, {Doyon}, {Malo}
   \& {Artigau}}{{Gagn{\'e}} et~al.}{2015b}]{gagne15a}
{Gagn{\'e}} J.,  {Lafreni{\`e}re} D.,  {Doyon} R.,  {Malo} L.,   {Artigau}
  {\'E}.,  2015b, \mn@doi [ApJ] {10.1088/0004-637X/798/2/73}, \href
  {http://cdsads.u-strasbg.fr/abs/2015ApJ...798...73G} {798, 73}

\bibitem[\protect\citeauthoryear{{Gagn{\'e}} et~al.,}{{Gagn{\'e}}
  et~al.}{2017a}]{gagne17a}
{Gagn{\'e}} J.,  et~al., 2017a, \mn@doi [ApJS] {10.3847/1538-4365/228/2/18},
  \href {http://cdsads.u-strasbg.fr/abs/2017ApJS..228...18G} {228, 18}

\bibitem[\protect\citeauthoryear{{Gagn{\'e}} et~al.,}{{Gagn{\'e}}
  et~al.}{2017b}]{gagne17b}
{Gagn{\'e}} J.,  et~al., 2017b, \mn@doi [ApJL] {10.3847/2041-8213/aa70e2},
  \href {http://cdsads.u-strasbg.fr/abs/2017ApJ...841L...1G} {841, L1}

\bibitem[\protect\citeauthoryear{{Gaia Collaboration} et~al.,}{{Gaia
  Collaboration} et~al.}{2018}]{Gaia_Brown2018}
{Gaia Collaboration} et~al., 2018, \mn@doi [A\&A]
  {10.1051/0004-6361/201833051}, \href
  {https://ui.adsabs.harvard.edu/abs/2018A&A...616A...1G} {616, A1}

\bibitem[\protect\citeauthoryear{{Gauza}, {B{\'e}jar}, {P{\'e}rez-Garrido},
  {Rosa Zapatero Osorio}, {Lodieu}, {Rebolo}, {Pall{\'e}}  \& {Nowak}}{{Gauza}
  et~al.}{2015}]{gauza15a}
{Gauza} B.,  {B{\'e}jar} V.~J.~S.,  {P{\'e}rez-Garrido} A.,  {Rosa Zapatero
  Osorio} M.,  {Lodieu} N.,  {Rebolo} R.,  {Pall{\'e}} E.,   {Nowak} G.,  2015,
  \mn@doi [ApJ] {10.1088/0004-637X/804/2/96}, \href
  {http://cdsads.u-strasbg.fr/abs/2015ApJ...804...96G} {804, 96}

\bibitem[\protect\citeauthoryear{{Geballe} et~al.,}{{Geballe}
  et~al.}{2002}]{geballe02}
{Geballe} T.~R.,  et~al., 2002, ApJ, \href
  {http://cdsads.u-strasbg.fr/cgi-bin/nph-bib_query?bibcode=2002ApJ...564..466G&amp;db_key=AST}
  {564, 466}

\bibitem[\protect\citeauthoryear{{Gorlova}, {Meyer}, {Rieke}  \&
  {Liebert}}{{Gorlova} et~al.}{2003}]{gorlova03}
{Gorlova} N.~I.,  {Meyer} M.~R.,  {Rieke} G.~H.,   {Liebert} J.,  2003, ApJ,
  \href
  {http://cdsads.u-strasbg.fr/cgi-bin/nph-bib_query?bibcode=2003ApJ...593.1074G&amp;db_key=AST}
  {593, 1074}

\bibitem[\protect\citeauthoryear{{Greenfield} \& {White}}{{Greenfield} \&
  {White}}{2006}]{greenfield06a}
{Greenfield} P.,  {White} R.~L.,  2006, in {Koekemoer} A.~M.,  {Goudfrooij} P.,
    {Dressel} L.~L.,  eds, The 2005 HST Calibration Workshop: Hubble After the
  Transition to Two-Gyro Mode. p.~437

\bibitem[\protect\citeauthoryear{{Haisch}, {Barsony}  \& {Tinney}}{{Haisch}
  et~al.}{2010}]{haisch10}
{Haisch} Jr. K.~E.,  {Barsony} M.,   {Tinney} C.,  2010, \mn@doi [ApJL]
  {10.1088/2041-8205/719/1/L90}, \href
  {http://cdsads.u-strasbg.fr/abs/2010ApJ...719L..90H} {719, L90}

\bibitem[\protect\citeauthoryear{{Hambly} et~al.,}{{Hambly}
  et~al.}{2008}]{hambly08}
{Hambly} N.~C.,  et~al., 2008, \mn@doi [MNRAS]
  {10.1111/j.1365-2966.2007.12700.x}, \href
  {http://cdsads.u-strasbg.fr/abs/2008MNRAS.384..637H} {384, 637}

\bibitem[\protect\citeauthoryear{{Hewett}, {Warren}, {Leggett}  \&
  {Hodgkin}}{{Hewett} et~al.}{2006}]{hewett06}
{Hewett} P.~C.,  {Warren} S.~J.,  {Leggett} S.~K.,   {Hodgkin} S.~T.,  2006,
  \mn@doi [MNRAS] {10.1111/j.1365-2966.2005.09969.x}, \href
  {http://cdsads.u-strasbg.fr/cgi-bin/nph-bib_query?bibcode=2006MNRAS.367..454H&db_key=AST}
  {367, 454}

\bibitem[\protect\citeauthoryear{{Hodgkin}, {Irwin}, {Hewett}  \&
  {Warren}}{{Hodgkin} et~al.}{2009}]{hodgkin09}
{Hodgkin} S.~T.,  {Irwin} M.~J.,  {Hewett} P.~C.,   {Warren} S.~J.,  2009,
  \mn@doi [MNRAS] {10.1111/j.1365-2966.2008.14387.x}, \href
  {http://cdsads.u-strasbg.fr/abs/2009MNRAS.394..675H} {394, 675}

\bibitem[\protect\citeauthoryear{{Irwin} et~al.,}{{Irwin}
  et~al.}{2004}]{irwin04}
{Irwin} M.~J.,  et~al., 2004, in {Quinn} P.~J.,  {Bridger} A.,  eds, Optimizing
  Scientific Return for Astronomy through Information Technologies. Edited by
  Quinn, Peter J.; Bridger, Alan. Proceedings of the SPIE, Volume 5493, pp.
  411-422 (2004).. pp 411--422, \mn@doi{10.1117/12.551449}

\bibitem[\protect\citeauthoryear{{Iye} et~al.,}{{Iye} et~al.}{2004}]{iye04a}
{Iye} M.,  et~al., 2004, \mn@doi [PASJ] {10.1093/pasj/56.2.381}, \href
  {https://ui.adsabs.harvard.edu/abs/2004PASJ...56..381I} {56, 381}

\bibitem[\protect\citeauthoryear{{Jenness} \& {LSST Data Management
  Team}}{{Jenness} \& {LSST Data Management Team}}{2017}]{jenness2017}
{Jenness} T.,  {LSST Data Management Team} 2017, in {Lorente} N.~P.~F.,
  {Shortridge} K.,   {Wayth} R.,  eds,  Astronomical Society of the Pacific
  Conference Series Vol. 512, Astronomical Data Analysis Software and Systems
  XXV. p.~297 (\mn@eprint {arXiv} {1511.06790})

\bibitem[\protect\citeauthoryear{{Kausch} et~al.,}{{Kausch}
  et~al.}{2015}]{kausch15}
{Kausch} W.,  et~al., 2015, \mn@doi [A\&A] {10.1051/0004-6361/201423909}, \href
  {http://cdsads.u-strasbg.fr/abs/2015A%26A...576A..78K} {576, A78}

\bibitem[\protect\citeauthoryear{{Kawanomoto} et~al.,}{{Kawanomoto}
  et~al.}{2018}]{kawanomoto18a}
{Kawanomoto} S.,  et~al., 2018, \mn@doi [PASJ] {10.1093/pasj/psy056}, \href
  {https://ui.adsabs.harvard.edu/abs/2018PASJ...70...66K} {70, 66}

\bibitem[\protect\citeauthoryear{{Kirkpatrick}}{{Kirkpatrick}}{2005}]{kirkpatrick05}
{Kirkpatrick} J.~D.,  2005, ARA\&A, \href
  {http://cdsads.u-strasbg.fr/cgi-bin/nph-bib_query?bibcode=2005ARA%26A..43..195K&db_key=AST}
  {43, 195}

\bibitem[\protect\citeauthoryear{{Kirkpatrick} et~al.,}{{Kirkpatrick}
  et~al.}{2019}]{kirkpatrick19}
{Kirkpatrick} J.~D.,  et~al., 2019, \mn@doi [ApJS] {10.3847/1538-4365/aaf6af},
  \href {https://ui.adsabs.harvard.edu/abs/2019ApJS..240...19K} {240, 19}

\bibitem[\protect\citeauthoryear{{Komiyama} et~al.,}{{Komiyama}
  et~al.}{2018}]{komiyama18a}
{Komiyama} Y.,  et~al., 2018, \mn@doi [PASJ] {10.1093/pasj/psx069}, \href
  {https://ui.adsabs.harvard.edu/abs/2018PASJ...70S...2K} {70, S2}

\bibitem[\protect\citeauthoryear{{Kratter}, {Murray-Clay}  \&
  {Youdin}}{{Kratter} et~al.}{2010}]{kratter10b}
{Kratter} K.~M.,  {Murray-Clay} R.~A.,   {Youdin} A.~N.,  2010, \mn@doi [ApJ]
  {10.1088/0004-637X/710/2/1375}, \href
  {http://cdsads.u-strasbg.fr/abs/2010ApJ...710.1375K} {710, 1375}

\bibitem[\protect\citeauthoryear{{Kraus}, {Ireland}, {Martinache}  \&
  {Lloyd}}{{Kraus} et~al.}{2008}]{kraus08a}
{Kraus} A.~L.,  {Ireland} M.~J.,  {Martinache} F.,   {Lloyd} J.~P.,  2008,
  \mn@doi [ApJ] {10.1086/587435}, \href
  {http://cdsads.u-strasbg.fr/abs/2008ApJ...679..762K} {679, 762}

\bibitem[\protect\citeauthoryear{{Kraus}, {Cody}, {Covey}, {Rizzuto}, {Mann}
  \& {Ireland}}{{Kraus} et~al.}{2015}]{kraus15a}
{Kraus} A.~L.,  {Cody} A.~M.,  {Covey} K.~R.,  {Rizzuto} A.~C.,  {Mann} A.~W.,
   {Ireland} M.~J.,  2015, \mn@doi [ApJ] {10.1088/0004-637X/807/1/3}, \href
  {http://cdsads.u-strasbg.fr/abs/2015ApJ...807....3K} {807, 3}

\bibitem[\protect\citeauthoryear{{Kunkel}}{{Kunkel}}{1999}]{kunkel99}
{Kunkel} M.,  1999, Ph.D.~Thesis, Julius-Maximilians-Universit\"at W\"urzburg,
  \href
  {http://cdsads.u-strasbg.fr/cgi-bin/nph-bib_query?bibcode=1998PhDT........14A&amp;db_key=AST}
  {}

\bibitem[\protect\citeauthoryear{{Lafreni{\`e}re}, {Jayawardhana}  \& {van
  Kerkwijk}}{{Lafreni{\`e}re} et~al.}{2010}]{lafreniere10a}
{Lafreni{\`e}re} D.,  {Jayawardhana} R.,   {van Kerkwijk} M.~H.,  2010, \mn@doi
  [ApJ] {10.1088/0004-637X/719/1/497}, \href
  {http://cdsads.u-strasbg.fr/abs/2010ApJ...719..497L} {719, 497}

\bibitem[\protect\citeauthoryear{{Lafreni{\`e}re}, {Jayawardhana}, {Janson},
  {Helling}, {Witte}  \& {Hauschildt}}{{Lafreni{\`e}re}
  et~al.}{2011}]{lafreniere11}
{Lafreni{\`e}re} D.,  {Jayawardhana} R.,  {Janson} M.,  {Helling} C.,  {Witte}
  S.,   {Hauschildt} P.,  2011, \mn@doi [ApJ] {10.1088/0004-637X/730/1/42},
  \href {http://cdsads.u-strasbg.fr/abs/2011ApJ...730...42L} {730, 42}

\bibitem[\protect\citeauthoryear{{Lafreni{\`e}re}, {Jayawardhana}, {van
  Kerkwijk}, {Brandeker}  \& {Janson}}{{Lafreni{\`e}re}
  et~al.}{2014}]{lafreniere14}
{Lafreni{\`e}re} D.,  {Jayawardhana} R.,  {van Kerkwijk} M.~H.,  {Brandeker}
  A.,   {Janson} M.,  2014, \mn@doi [ApJ] {10.1088/0004-637X/785/1/47}, \href
  {http://cdsads.u-strasbg.fr/abs/2014ApJ...785...47L} {785, 47}

\bibitem[\protect\citeauthoryear{{Lawrence}, {Warren}, {Almaini}, {Edge},
  {Hambly}  \& {17 co-authors}}{{Lawrence} et~al.}{2007}]{lawrence07}
{Lawrence} A.,  {Warren} S.~J.,  {Almaini} O.,  {Edge} A.~C.,  {Hambly} N.~C.,
   {17 co-authors} 2007, \mn@doi [MNRAS] {10.1111/j.1365-2966.2007.12040.x},
  \href {http://cdsads.u-strasbg.fr/abs/2007MNRAS.379.1599L} {379, 1599}

\bibitem[\protect\citeauthoryear{{Leggett} et~al.,}{{Leggett}
  et~al.}{2000}]{leggett00}
{Leggett} S.~K.,  et~al., 2000, ApJL, \href
  {http://cdsads.u-strasbg.fr/cgi-bin/nph-bib_query?bibcode=2000ApJ...536L..35L&amp;db_key=AST}
  {536, L35}

\bibitem[\protect\citeauthoryear{{Lissauer}, {Dawson}  \&
  {Tremaine}}{{Lissauer} et~al.}{2014}]{lissauer14}
{Lissauer} J.~J.,  {Dawson} R.~I.,   {Tremaine} S.,  2014, \mn@doi [Nature]
  {10.1038/nature13781}, \href
  {http://adsabs.harvard.edu/abs/2014Natur.513..336L} {513, 336}

\bibitem[\protect\citeauthoryear{{Liu} et~al.,}{{Liu} et~al.}{2013}]{liu13a}
{Liu} M.~C.,  et~al., 2013, \mn@doi [ApJL] {10.1088/2041-8205/777/2/L20}, \href
  {http://cdsads.u-strasbg.fr/abs/2013ApJ...777L..20L} {777, L20}

\bibitem[\protect\citeauthoryear{{Liu}, {Dupuy}  \& {Allers}}{{Liu}
  et~al.}{2016}]{liu16a}
{Liu} M.~C.,  {Dupuy} T.~J.,   {Allers} K.~N.,  2016, \mn@doi [ApJ]
  {10.3847/1538-4357/833/1/96}, \href
  {http://cdsads.u-strasbg.fr/abs/2016ApJ...833...96L} {833, 96}

\bibitem[\protect\citeauthoryear{{Lodieu}}{{Lodieu}}{2013}]{lodieu13c}
{Lodieu} N.,  2013, \mn@doi [MNRAS] {10.1093/mnras/stt402}, \href
  {http://cdsads.u-strasbg.fr/abs/2013MNRAS.431.3222L} {431, 3222}

\bibitem[\protect\citeauthoryear{{Lodieu}, {Hambly}  \& {Jameson}}{{Lodieu}
  et~al.}{2006}]{lodieu06}
{Lodieu} N.,  {Hambly} N.~C.,   {Jameson} R.~F.,  2006, \mn@doi [MNRAS]
  {10.1111/j.1365-2966.2006.10958.x}, \href
  {http://cdsads.u-strasbg.fr/cgi-bin/nph-bib_query?bibcode=2006MNRAS.373...95L&db_key=AST}
  {373, 95}

\bibitem[\protect\citeauthoryear{{Lodieu}, {Hambly}, {Jameson}, {Hodgkin},
  {Carraro}  \& {Kendall}}{{Lodieu} et~al.}{2007a}]{lodieu07a}
{Lodieu} N.,  {Hambly} N.~C.,  {Jameson} R.~F.,  {Hodgkin} S.~T.,  {Carraro}
  G.,   {Kendall} T.~R.,  2007a, \mn@doi [MNRAS]
  {10.1111/j.1365-2966.2006.11151.x}, \href
  {http://cdsads.u-strasbg.fr/cgi-bin/nph-bib_query?bibcode=2007MNRAS.374..372L&db_key=AST}
  {374, 372}

\bibitem[\protect\citeauthoryear{{Lodieu}, {Pinfield}, {Leggett}, {Jameson},
  {Mortlock}, {Warren}  \& {co-authors}}{{Lodieu} et~al.}{2007b}]{lodieu07b}
{Lodieu} N.,  {Pinfield} D.~J.,  {Leggett} S.~K.,  {Jameson} R.~F.,  {Mortlock}
  D.~J.,  {Warren} S.~J.,   {co-authors} .,  2007b, \mn@doi [MNRAS]
  {10.1111/j.1365-2966.2007.12023.x}, \href
  {http://cdsads.u-strasbg.fr/abs/2007MNRAS.379.1423L} {379, 1423}

\bibitem[\protect\citeauthoryear{{Lodieu}, {Hambly}, {Jameson}  \&
  {Hodgkin}}{{Lodieu} et~al.}{2008}]{lodieu08a}
{Lodieu} N.,  {Hambly} N.~C.,  {Jameson} R.~F.,   {Hodgkin} S.~T.,  2008,
  \mn@doi [MNRAS] {10.1111/j.1365-2966.2007.12676.x}, \href
  {http://adsabs.harvard.edu/abs/2008MNRAS.383.1385L} {383, 1385}

\bibitem[\protect\citeauthoryear{{Lodieu}, {Hambly}, {Dobbie}, {Cross},
  {Christensen}, {Martin}  \& {Valdivielso}}{{Lodieu}
  et~al.}{2011a}]{lodieu11c}
{Lodieu} N.,  {Hambly} N.~C.,  {Dobbie} P.~D.,  {Cross} N.~J.~G.,
  {Christensen} L.,  {Martin} E.~L.,   {Valdivielso} L.,  2011a, \mn@doi
  [MNRAS] {10.1111/j.1365-2966.2011.19651.x}, \href
  {http://cdsads.u-strasbg.fr/abs/2011MNRAS.418.2604L} {418, 2604}

\bibitem[\protect\citeauthoryear{{Lodieu}, {Dobbie}  \& {Hambly}}{{Lodieu}
  et~al.}{2011b}]{lodieu11a}
{Lodieu} N.,  {Dobbie} P.~D.,   {Hambly} N.~C.,  2011b, \mn@doi [A\&A]
  {10.1051/0004-6361/201014992}, \href
  {http://cdsads.u-strasbg.fr/abs/2011A%26A...527A..24L} {527, A24}

\bibitem[\protect\citeauthoryear{{Lodieu}, {Ivanov}  \& {Dobbie}}{{Lodieu}
  et~al.}{2013a}]{lodieu13b}
{Lodieu} N.,  {Ivanov} V.~D.,   {Dobbie} P.~D.,  2013a, \mn@doi [MNRAS]
  {10.1093/mnras/sts726}, \href
  {http://cdsads.u-strasbg.fr/abs/2013MNRAS.430.1784L} {430, 1784}

\bibitem[\protect\citeauthoryear{{Lodieu}, {Dobbie}, {Cross}, {Hambly}, {Read},
  {Blake}  \& {Floyd}}{{Lodieu} et~al.}{2013b}]{lodieu13d}
{Lodieu} N.,  {Dobbie} P.~D.,  {Cross} N.~J.~G.,  {Hambly} N.~C.,  {Read}
  M.~A.,  {Blake} R.~P.,   {Floyd} D.~J.~E.,  2013b, \mn@doi [MNRAS]
  {10.1093/mnras/stt1460}, \href
  {http://cdsads.u-strasbg.fr/abs/2013MNRAS.435.2474L} {435, 2474}

\bibitem[\protect\citeauthoryear{{Lodieu} et~al.,}{{Lodieu}
  et~al.}{2015}]{lodieu15c}
{Lodieu} N.,  et~al., 2015, \mn@doi [A\&A] {10.1051/0004-6361/201527464}, \href
  {http://cdsads.u-strasbg.fr/abs/2015A%26A...584A.128L} {584, A128}

\bibitem[\protect\citeauthoryear{{Lodieu}, {Zapatero Osorio}, {B{\'e}jar}  \&
  {Pe{\~n}a Ram{\'{\i}}rez}}{{Lodieu} et~al.}{2018a}]{lodieu18a}
{Lodieu} N.,  {Zapatero Osorio} M.~R.,  {B{\'e}jar} V.~J.~S.,   {Pe{\~n}a
  Ram{\'{\i}}rez} K.,  2018a, \mn@doi [MNRAS] {10.1093/mnras/stx2279}, \href
  {http://cdsads.u-strasbg.fr/abs/2018MNRAS.473.2020L} {473, 2020}

\bibitem[\protect\citeauthoryear{{Lodieu}, {Rebolo}  \&
  {P{\'e}rez-Garrido}}{{Lodieu} et~al.}{2018b}]{lodieu18b}
{Lodieu} N.,  {Rebolo} R.,   {P{\'e}rez-Garrido} A.,  2018b, \mn@doi [A\&A]
  {10.1051/0004-6361/201832748}, \href
  {http://cdsads.u-strasbg.fr/abs/2018A%26A...615L..12L} {615, L12}

\bibitem[\protect\citeauthoryear{{Lodieu}, {Paunzen}  \& {Zejda}}{{Lodieu}
  et~al.}{2020}]{lodieu20a}
{Lodieu} N.,  {Paunzen} E.,   {Zejda} M.,  2020, {Low-Mass and Sub-stellar
  Eclipsing Binaries in Stellar Clusters}.
pp 213--243, \mn@doi{10.1007/978-3-030-38509-5_8}

\bibitem[\protect\citeauthoryear{{Looper}, {Kirkpatrick}  \&
  {Burgasser}}{{Looper} et~al.}{2007}]{looper07}
{Looper} D.~L.,  {Kirkpatrick} J.~D.,   {Burgasser} A.~J.,  2007, \mn@doi [AJ]
  {10.1086/520645}, \href {http://adsabs.harvard.edu/abs/2007AJ....134.1162L}
  {134, 1162}

\bibitem[\protect\citeauthoryear{{Low} \& {Lynden-Bell}}{{Low} \&
  {Lynden-Bell}}{1976}]{low76}
{Low} C.,  {Lynden-Bell} D.,  1976, MNRAS, \href
  {http://cdsads.u-strasbg.fr/cgi-bin/nph-bib_query?bibcode=1976MNRAS.176..367L&amp;db_key=AST}
  {176, 367}

\bibitem[\protect\citeauthoryear{{Lucas} \& {Roche}}{{Lucas} \&
  {Roche}}{2000}]{lucas00}
{Lucas} P.~W.,  {Roche} P.~F.,  2000, MNRAS, \href
  {http://cdsads.u-strasbg.fr/cgi-bin/nph-bib_query?bibcode=2000MNRAS.314..858L&amp;db_key=AST}
  {314, 858}

\bibitem[\protect\citeauthoryear{{Luhman}}{{Luhman}}{2012}]{luhman12b}
{Luhman} K.~L.,  2012, \mn@doi [ARA\&A] {10.1146/annurev-astro-081811-125528},
  \href {http://cdsads.u-strasbg.fr/abs/2012ARA%26A..50...65L} {50, 65}

\bibitem[\protect\citeauthoryear{{Luhman} \& {Esplin}}{{Luhman} \&
  {Esplin}}{2020a}]{luhman20}
{Luhman} K.~L.,  {Esplin} T.~L.,  2020a, \mn@doi [AJ]
  {10.3847/1538-3881/ab9599}, \href
  {https://ui.adsabs.harvard.edu/abs/2020AJ....160...44L} {160, 44}

\bibitem[\protect\citeauthoryear{{Luhman} \& {Esplin}}{{Luhman} \&
  {Esplin}}{2020b}]{luhman20a}
{Luhman} K.~L.,  {Esplin} T.~L.,  2020b, \mn@doi [AJ]
  {10.3847/1538-3881/ab9599}, \href
  {https://ui.adsabs.harvard.edu/abs/2020AJ....160...44L} {160, 44}

\bibitem[\protect\citeauthoryear{{Luhman} \& {Mamajek}}{{Luhman} \&
  {Mamajek}}{2012}]{luhman12c}
{Luhman} K.~L.,  {Mamajek} E.~E.,  2012, \mn@doi [ApJ]
  {10.1088/0004-637X/758/1/31}, \href
  {http://cdsads.u-strasbg.fr/abs/2012ApJ...758...31L} {758, 31}

\bibitem[\protect\citeauthoryear{{Luhman}, {Rieke}, {Lada}  \& {Lada}}{{Luhman}
  et~al.}{1998}]{luhman98}
{Luhman} K.~L.,  {Rieke} G.~H.,  {Lada} C.~J.,   {Lada} E.~A.,  1998, ApJ,
  \href
  {http://cdsads.u-strasbg.fr/cgi-bin/nph-bib_query?bibcode=1998ApJ...508..347L&db_key=AST}
  {508, 347}

\bibitem[\protect\citeauthoryear{{Luhman}, {Herrmann}, {Mamajek}, {Esplin}  \&
  {Pecaut}}{{Luhman} et~al.}{2018}]{luhman18a}
{Luhman} K.~L.,  {Herrmann} K.~A.,  {Mamajek} E.~E.,  {Esplin} T.~L.,
  {Pecaut} M.~J.,  2018, \mn@doi [AJ] {10.3847/1538-3881/aacc6d}, \href
  {https://ui.adsabs.harvard.edu/abs/2018AJ....156...76L} {156, 76}

\bibitem[\protect\citeauthoryear{{Manjavacas} et~al.,}{{Manjavacas}
  et~al.}{2014}]{manjavacas14}
{Manjavacas} E.,  et~al., 2014, \mn@doi [A\&A] {10.1051/0004-6361/201323016},
  \href {http://cdsads.u-strasbg.fr/abs/2014A%26A...564A..55M} {564, A55}

\bibitem[\protect\citeauthoryear{{Mann} et~al.,}{{Mann} et~al.}{2016}]{mann16b}
{Mann} A.~W.,  et~al., 2016, \mn@doi [AJ] {10.3847/0004-6256/152/3/61}, \href
  {http://cdsads.u-strasbg.fr/abs/2016AJ....152...61M} {152, 61}

\bibitem[\protect\citeauthoryear{{Marocco} et~al.,}{{Marocco}
  et~al.}{2014}]{marocco14}
{Marocco} F.,  et~al., 2014, \mn@doi [MNRAS] {10.1093/mnras/stt2463}, \href
  {https://ui.adsabs.harvard.edu/abs/2014MNRAS.439..372M} {439, 372}

\bibitem[\protect\citeauthoryear{{Mart\'{\i}n}, {Rebolo}  \& {Zapatero
  Osorio}}{{Mart\'{\i}n} et~al.}{1996}]{martin96}
{Mart\'{\i}n} E.~L.,  {Rebolo} R.,   {Zapatero Osorio} M.~R.,  1996, ApJ, \href
  {http://cdsads.u-strasbg.fr/cgi-bin/nph-bib_query?bibcode=1996ApJ...469..706M&amp;db_key=AST}
  {469, 706}

\bibitem[\protect\citeauthoryear{{Mart{\'{\i}}n}, {Delfosse}  \&
  {Guieu}}{{Mart{\'{\i}}n} et~al.}{2004}]{martin04}
{Mart{\'{\i}}n} E.~L.,  {Delfosse} X.,   {Guieu} S.,  2004, AJ, \href
  {http://cdsads.u-strasbg.fr/cgi-bin/nph-bib_query?bibcode=2004AJ....127..449M&amp;db_key=AST}
  {127, 449}

\bibitem[\protect\citeauthoryear{{Mauro}, {Moni Bidin}, {Chen{\'e}}, {Geisler},
  {Alonso-Garc{\'{\i}}a}, {Borissova}  \& {Carraro}}{{Mauro}
  et~al.}{2013}]{mauro13a}
{Mauro} F.,  {Moni Bidin} C.,  {Chen{\'e}} A.-N.,  {Geisler} D.,
  {Alonso-Garc{\'{\i}}a} J.,  {Borissova} J.,   {Carraro} G.,  2013, Revista
  Mexicana de Astronom\'{\i}a y Astrof\'{\i}sica, \href
  {http://adsabs.harvard.edu/abs/2013RMxAA..49..189M} {49, 189}

\bibitem[\protect\citeauthoryear{{Mayor} \& {Queloz}}{{Mayor} \&
  {Queloz}}{1995}]{mayor95}
{Mayor} M.,  {Queloz} D.,  1995, Nature, \href
  {http://cdsads.u-strasbg.fr/cgi-bin/nph-bib_query?bibcode=1995Natur.378..355M&db_key=AST}
  {378, 355}

\bibitem[\protect\citeauthoryear{{McGovern}, {Kirkpatrick}, {McLean},
  {Burgasser}, {Prato}  \& {Lowrance}}{{McGovern} et~al.}{2004}]{mcgovern04}
{McGovern} M.~R.,  {Kirkpatrick} J.~D.,  {McLean} I.~S.,  {Burgasser} A.~J.,
  {Prato} L.,   {Lowrance} P.~J.,  2004, \mn@doi [ApJ] {10.1086/379849}, \href
  {http://cdsads.u-strasbg.fr/cgi-bin/nph-bib_query?bibcode=2004ApJ...600.1020M&db_key=AST}
  {600, 1020}

\bibitem[\protect\citeauthoryear{{Miller} \& {Scalo}}{{Miller} \&
  {Scalo}}{1979}]{miller79}
{Miller} G.~E.,  {Scalo} J.~M.,  1979, ApJS, \href
  {http://cdsads.u-strasbg.fr/cgi-bin/nph-bib_query?bibcode=1979ApJS...41..513M&db_key=AST}
  {41, 513}

\bibitem[\protect\citeauthoryear{{Miyazaki} et~al.,}{{Miyazaki}
  et~al.}{2018}]{miyazaki18a}
{Miyazaki} S.,  et~al., 2018, \mn@doi [PASJ] {10.1093/pasj/psx063}, \href
  {https://ui.adsabs.harvard.edu/abs/2018PASJ...70S...1M} {70, S1}

\bibitem[\protect\citeauthoryear{{Mu{\v z}i{\'c}}, {Scholz}, {Geers},
  {Jayawardhana}  \& {L{\'o}pez Mart{\'{\i}}}}{{Mu{\v z}i{\'c}}
  et~al.}{2014}]{muzic14}
{Mu{\v z}i{\'c}} K.,  {Scholz} A.,  {Geers} V.~C.,  {Jayawardhana} R.,
  {L{\'o}pez Mart{\'{\i}}} B.,  2014, \mn@doi [ApJ]
  {10.1088/0004-637X/785/2/159}, \href
  {http://cdsads.u-strasbg.fr/abs/2014ApJ...785..159M} {785, 159}

\bibitem[\protect\citeauthoryear{{Nakajima}, {Oppenheimer}, {Kulkarni},
  {Golimowski}, {Matthews}  \& {Durrance}}{{Nakajima}
  et~al.}{1995}]{nakajima95}
{Nakajima} T.,  {Oppenheimer} B.~R.,  {Kulkarni} S.~R.,  {Golimowski} D.~A.,
  {Matthews} K.,   {Durrance} S.~T.,  1995, Nature, \href
  {http://cdsads.u-strasbg.fr/cgi-bin/nph-bib_query?bibcode=1995Natur.378..463N&amp;db_key=AST}
  {378, 463}

\bibitem[\protect\citeauthoryear{{Ochsenbein}, {Bauer}  \&
  {Marcout}}{{Ochsenbein} et~al.}{2000}]{ochsenbein00}
{Ochsenbein} F.,  {Bauer} P.,   {Marcout} J.,  2000, \mn@doi [A\&AS]
  {10.1051/aas:2000169}, \href
  {http://cdsads.u-strasbg.fr/abs/2000A%26AS..143...23O} {143, 23}

\bibitem[\protect\citeauthoryear{{Pe{\~n}a Ram{\'{\i}}rez}, {Zapatero Osorio},
  {B{\'e}jar}, {Rebolo}  \& {Bihain}}{{Pe{\~n}a Ram{\'{\i}}rez}
  et~al.}{2011}]{penya11a}
{Pe{\~n}a Ram{\'{\i}}rez} K.,  {Zapatero Osorio} M.~R.,  {B{\'e}jar} V.~J.~S.,
  {Rebolo} R.,   {Bihain} G.,  2011, \mn@doi [A\&A]
  {10.1051/0004-6361/201116812}, \href
  {http://cdsads.u-strasbg.fr/abs/2011A%26A...532A..42P} {532, A42}

\bibitem[\protect\citeauthoryear{{Pe{\~n}a Ram{\'{\i}}rez}, {Zapatero Osorio}
  \& {B{\'e}jar}}{{Pe{\~n}a Ram{\'{\i}}rez} et~al.}{2015}]{penya15}
{Pe{\~n}a Ram{\'{\i}}rez} K.,  {Zapatero Osorio} M.~R.,   {B{\'e}jar} V.~J.~S.,
   2015, \mn@doi [A\&A] {10.1051/0004-6361/201424816}, \href
  {http://cdsads.u-strasbg.fr/abs/2015A%26A...574A.118P} {574, A118}

\bibitem[\protect\citeauthoryear{{Pe{\~n}a Ram{\'{\i}}rez}, {B{\'e}jar}  \&
  {Zapatero Osorio}}{{Pe{\~n}a Ram{\'{\i}}rez} et~al.}{2016}]{penya16a}
{Pe{\~n}a Ram{\'{\i}}rez} K.,  {B{\'e}jar} V.~J.~S.,   {Zapatero Osorio} M.~R.,
   2016, \mn@doi [A\&A] {10.1051/0004-6361/201527425}, \href
  {http://cdsads.u-strasbg.fr/abs/2016A%26A...586A.157P} {586, A157}

\bibitem[\protect\citeauthoryear{{Pecaut}, {Mamajek}  \& {Bubar}}{{Pecaut}
  et~al.}{2012}]{pecaut12}
{Pecaut} M.~J.,  {Mamajek} E.~E.,   {Bubar} E.~J.,  2012, \mn@doi [ApJ]
  {10.1088/0004-637X/746/2/154}, \href
  {http://cdsads.u-strasbg.fr/abs/2012ApJ...746..154P} {746, 154}

\bibitem[\protect\citeauthoryear{{Pinfield}, {Burningham}, {Tamura}, {Leggett},
  {Lodieu}, {Lucas}, {Mortlock}  \& {28 co-authors}}{{Pinfield}
  et~al.}{2008}]{pinfield08}
{Pinfield} D.~J.,  {Burningham} B.,  {Tamura} M.,  {Leggett} S.~K.,  {Lodieu}
  N.,  {Lucas} P.~W.,  {Mortlock} D.~J.,   {28 co-authors} 2008, \mn@doi
  [MNRAS] {10.1111/j.1365-2966.2008.13729.x}, \href
  {http://cdsads.u-strasbg.fr/abs/2008MNRAS.390..304P} {390, 304}

\bibitem[\protect\citeauthoryear{{Preibisch} \& {Zinnecker}}{{Preibisch} \&
  {Zinnecker}}{1999}]{preibisch99}
{Preibisch} T.,  {Zinnecker} H.,  1999, \mn@doi [AJ] {10.1086/300842}, \href
  {http://cdsads.u-strasbg.fr/cgi-bin/nph-bib_query?bibcode=1999AJ....117.2381P&db_key=AST}
  {117, 2381}

\bibitem[\protect\citeauthoryear{{Preibisch} \& {Zinnecker}}{{Preibisch} \&
  {Zinnecker}}{2002}]{preibisch02}
{Preibisch} T.,  {Zinnecker} H.,  2002, AJ, \href
  {http://cdsads.u-strasbg.fr/cgi-bin/nph-bib_query?bibcode=2002AJ....123.1613P&amp;db_key=AST}
  {123, 1613}

\bibitem[\protect\citeauthoryear{{Preibisch}, {Guenther}, {Zinnecker},
  {Sterzik}, {Frink}  \& {Roeser}}{{Preibisch} et~al.}{1998}]{preibisch98}
{Preibisch} T.,  {Guenther} E.,  {Zinnecker} H.,  {Sterzik} M.,  {Frink} S.,
  {Roeser} S.,  1998, A\&A, \href
  {http://cdsads.u-strasbg.fr/cgi-bin/nph-bib_query?bibcode=1998A%26A...333..619P&db_key=AST}
  {333, 619}

\bibitem[\protect\citeauthoryear{{Preibisch}, {Guenther}  \&
  {Zinnecker}}{{Preibisch} et~al.}{2001}]{preibisch01}
{Preibisch} T.,  {Guenther} E.,   {Zinnecker} H.,  2001, \mn@doi [AJ]
  {10.1086/318774}, \href
  {http://cdsads.u-strasbg.fr/cgi-bin/nph-bib_query?bibcode=2001AJ....121.1040P&db_key=AST}
  {121, 1040}

\bibitem[\protect\citeauthoryear{{Rebolo}, {Zapatero-Osorio}  \&
  {Mart\'{\i}n}}{{Rebolo} et~al.}{1995}]{rebolo95}
{Rebolo} R.,  {Zapatero-Osorio} M.~R.,   {Mart\'{\i}n} E.~L.,  1995, Nature,
  \href
  {http://cdsads.u-strasbg.fr/cgi-bin/nph-bib_query?bibcode=1995Natur.377..129R&amp;db_key=AST}
  {377, 129}

\bibitem[\protect\citeauthoryear{{Rees}}{{Rees}}{1976}]{rees76}
{Rees} M.~J.,  1976, MNRAS, \href
  {http://cdsads.u-strasbg.fr/cgi-bin/nph-bib_query?bibcode=1976MNRAS.176..483R&amp;db_key=AST}
  {176, 483}

\bibitem[\protect\citeauthoryear{{Rogers} \& {Wadsley}}{{Rogers} \&
  {Wadsley}}{2012}]{rogers12}
{Rogers} P.~D.,  {Wadsley} J.,  2012, \mn@doi [MNRAS]
  {10.1111/j.1365-2966.2012.21014.x}, \href
  {http://cdsads.u-strasbg.fr/abs/2012MNRAS.423.1896R} {423, 1896}

\bibitem[\protect\citeauthoryear{{Salpeter}}{{Salpeter}}{1955}]{salpeter55}
{Salpeter} E.~E.,  1955, ApJ, \href
  {http://cdsads.u-strasbg.fr/cgi-bin/nph-bib_query?bibcode=1955ApJ...121..161S&db_key=AST}
  {121, 161}

\bibitem[\protect\citeauthoryear{{Scalo}}{{Scalo}}{1986}]{scalo86}
{Scalo} J.~M.,  1986, Fundamentals of Cosmic Physics, \href
  {http://cdsads.u-strasbg.fr/cgi-bin/nph-bib_query?bibcode=1986FCPh...11....1S&db_key=AST}
  {11, 1}

\bibitem[\protect\citeauthoryear{{Schlafly}, {Meisner}  \& {Green}}{{Schlafly}
  et~al.}{2019}]{schlafly19}
{Schlafly} E.~F.,  {Meisner} A.~M.,   {Green} G.~M.,  2019, \mn@doi [ApJS]
  {10.3847/1538-4365/aafbea}, \href
  {https://ui.adsabs.harvard.edu/abs/2019ApJS..240...30S} {240, 30}

\bibitem[\protect\citeauthoryear{{Schmidt}, {West}, {Hawley}  \&
  {Pineda}}{{Schmidt} et~al.}{2010}]{schmidt10b}
{Schmidt} S.~J.,  {West} A.~A.,  {Hawley} S.~L.,   {Pineda} J.~S.,  2010,
  \mn@doi [AJ] {10.1088/0004-6256/139/5/1808}, \href
  {http://cdsads.u-strasbg.fr/abs/2010AJ....139.1808S} {139, 1808}

\bibitem[\protect\citeauthoryear{{Schneider}, {Windsor}, {Cushing},
  {Kirkpatrick}  \& {Wright}}{{Schneider} et~al.}{2016}]{schneider16b}
{Schneider} A.~C.,  {Windsor} J.,  {Cushing} M.~C.,  {Kirkpatrick} J.~D.,
  {Wright} E.~L.,  2016, \mn@doi [ApJL] {10.3847/2041-8205/822/1/L1}, \href
  {http://cdsads.u-strasbg.fr/abs/2016ApJ...822L...1S} {822, L1}

\bibitem[\protect\citeauthoryear{{Scholz}, {Muzic}, {Geers}, {Bonavita},
  {Jayawardhana}  \& {Tamura}}{{Scholz} et~al.}{2012}]{scholz12b}
{Scholz} A.,  {Muzic} K.,  {Geers} V.,  {Bonavita} M.,  {Jayawardhana} R.,
  {Tamura} M.,  2012, \mn@doi [ApJ] {10.1088/0004-637X/744/1/6}, \href
  {http://cdsads.u-strasbg.fr/abs/2012ApJ...744....6S} {744, 6}

\bibitem[\protect\citeauthoryear{{Science Software Branch at STScI}}{{Science
  Software Branch at STScI}}{2012}]{pyraf12a}
{Science Software Branch at STScI} 2012, PyRAF: Python alternative for IRAF,
  Astrophysics Source Code Library (\mn@eprint {ascl} {1207.011})

\bibitem[\protect\citeauthoryear{{Silk}}{{Silk}}{1977}]{silk77a}
{Silk} J.,  1977, \mn@doi [ApJ] {10.1086/155240}, \href
  {http://cdsads.u-strasbg.fr/abs/1977ApJ...214..152S} {214, 152}

\bibitem[\protect\citeauthoryear{{Slesnick}, {Carpenter}  \&
  {Hillenbrand}}{{Slesnick} et~al.}{2006}]{slesnick06}
{Slesnick} C.~L.,  {Carpenter} J.~M.,   {Hillenbrand} L.~A.,  2006, \mn@doi
  [AJ] {10.1086/503560}, \href
  {http://cdsads.u-strasbg.fr/cgi-bin/nph-bib_query?bibcode=2006AJ....131.3016S&db_key=AST}
  {131, 3016}

\bibitem[\protect\citeauthoryear{{Slesnick}, {Hillenbrand}  \&
  {Carpenter}}{{Slesnick} et~al.}{2008}]{slesnick08}
{Slesnick} C.~L.,  {Hillenbrand} L.~A.,   {Carpenter} J.~M.,  2008, \mn@doi
  [ApJ] {10.1086/592265}, \href
  {http://cdsads.u-strasbg.fr/abs/2008ApJ...688..377S} {688, 377}

\bibitem[\protect\citeauthoryear{{Smette} et~al.,}{{Smette}
  et~al.}{2015}]{smette15}
{Smette} A.,  et~al., 2015, \mn@doi [A\&A] {10.1051/0004-6361/201423932}, \href
  {http://cdsads.u-strasbg.fr/abs/2015A%26A...576A..77S} {576, A77}

\bibitem[\protect\citeauthoryear{{Song}, {Zuckerman}  \& {Bessell}}{{Song}
  et~al.}{2012}]{song12}
{Song} I.,  {Zuckerman} B.,   {Bessell} M.~S.,  2012, \mn@doi [AJ]
  {10.1088/0004-6256/144/1/8}, \href
  {http://cdsads.u-strasbg.fr/abs/2012AJ....144....8S} {144, 8}

\bibitem[\protect\citeauthoryear{{Spezzi}, {Alves de Oliveira}, {Moraux},
  {Bouvier}, {Winston}, {Hudelot}, {Bouy}  \& {Cuillandre}}{{Spezzi}
  et~al.}{2012}]{spezzi12b}
{Spezzi} L.,  {Alves de Oliveira} C.,  {Moraux} E.,  {Bouvier} J.,  {Winston}
  E.,  {Hudelot} P.,  {Bouy} H.,   {Cuillandre} J.-C.,  2012, \mn@doi [A\&A]
  {10.1051/0004-6361/201219559}, \href
  {http://cdsads.u-strasbg.fr/abs/2012A%26A...545A.105S} {545, A105}

\bibitem[\protect\citeauthoryear{{Stauffer}, {Schultz}  \&
  {Kirkpatrick}}{{Stauffer} et~al.}{1998}]{stauffer98}
{Stauffer} J.~R.,  {Schultz} G.,   {Kirkpatrick} J.~D.,  1998, ApJL, \href
  {http://cdsads.u-strasbg.fr/cgi-bin/nph-bib_query?bibcode=1998ApJ...499L.199S&db_key=AST}
  {499, 199}

\bibitem[\protect\citeauthoryear{{Stetson}}{{Stetson}}{1987}]{stetson87}
{Stetson} P.~B.,  1987, \mn@doi [PASP] {10.1086/131977}, \href
  {http://adsabs.harvard.edu/abs/1987PASP...99..191S} {99, 191}

\bibitem[\protect\citeauthoryear{{Tody}}{{Tody}}{1986}]{tody86}
{Tody} D.,  1986, in {Crawford} D.~L.,  ed.,  Society of Photo-Optical
  Instrumentation Engineers (SPIE) Conference Series Vol. 627, Society of
  Photo-Optical Instrumentation Engineers (SPIE) Conference Series. p.~733

\bibitem[\protect\citeauthoryear{{Tody}}{{Tody}}{1993}]{tody93}
{Tody} D.,  1993, in {Hanisch} R.~J.,  {Brissenden} R.~J.~V.,   {Barnes} J.,
  eds,  Astronomical Society of the Pacific Conference Series Vol. 52,
  Astronomical Data Analysis Software and Systems II. p.~173

\bibitem[\protect\citeauthoryear{{Vernet} et~al.,}{{Vernet}
  et~al.}{2011}]{vernet11}
{Vernet} J.,  et~al., 2011, \mn@doi [A\&A] {10.1051/0004-6361/201117752}, \href
  {http://cdsads.u-strasbg.fr/abs/2011A%26A...536A.105V} {536, A105}

\bibitem[\protect\citeauthoryear{{Walter}, {Vrba}, {Mathieu}, {Brown}  \&
  {Myers}}{{Walter} et~al.}{1994}]{walter94}
{Walter} F.~M.,  {Vrba} F.~J.,  {Mathieu} R.~D.,  {Brown} A.,   {Myers} P.~C.,
  1994, AJ, \href
  {http://cdsads.u-strasbg.fr/cgi-bin/nph-bib_query?bibcode=1994AJ....107..692W&amp;db_key=AST}
  {107, 692}

\bibitem[\protect\citeauthoryear{{Whitworth} \& {Stamatellos}}{{Whitworth} \&
  {Stamatellos}}{2006}]{whitworth06}
{Whitworth} A.~P.,  {Stamatellos} D.,  2006, \mn@doi [A\&A]
  {10.1051/0004-6361:20065806}, \href
  {http://cdsads.u-strasbg.fr/abs/2006A%26A...458..817W} {458, 817}

\bibitem[\protect\citeauthoryear{{Zapatero Osorio}, {B{\' e}jar},
  {Mart{\'{\i}}n}, {Rebolo}, {Barrado y Navascu{\' e}s}, {Bailer-Jones}  \&
  {Mundt}}{{Zapatero Osorio} et~al.}{2000}]{zapatero00}
{Zapatero Osorio} M.~R.,  {B{\' e}jar} V.~J.~S.,  {Mart{\'{\i}}n} E.~L.,
  {Rebolo} R.,  {Barrado y Navascu{\' e}s} D.,  {Bailer-Jones} C.~A.~L.,
  {Mundt} R.,  2000, Science, \href
  {http://cdsads.u-strasbg.fr/cgi-bin/nph-bib_query?bibcode=2000Sci...290..103Z&amp;db_key=AST}
  {290, 103}

\bibitem[\protect\citeauthoryear{{Zapatero Osorio}, {B{\'e}jar},
  {Miles-P{\'a}ez}, {Pe{\~n}a Ram{\'{\i}}rez}, {Rebolo}  \&
  {Pall{\'e}}}{{Zapatero Osorio} et~al.}{2014a}]{zapatero14a}
{Zapatero Osorio} M.~R.,  {B{\'e}jar} V.~J.~S.,  {Miles-P{\'a}ez} P.~A.,
  {Pe{\~n}a Ram{\'{\i}}rez} K.,  {Rebolo} R.,   {Pall{\'e}} E.,  2014a, \mn@doi
  [A\&A] {10.1051/0004-6361/201321340}, \href
  {http://cdsads.u-strasbg.fr/abs/2014A%26A...568A...6Z} {568, A6}

\bibitem[\protect\citeauthoryear{{Zapatero Osorio} et~al.,}{{Zapatero Osorio}
  et~al.}{2014b}]{zapatero14b}
{Zapatero Osorio} M.~R.,  et~al., 2014b, \mn@doi [A\&A]
  {10.1051/0004-6361/201423848}, \href
  {http://cdsads.u-strasbg.fr/abs/2014A%26A...568A..77Z} {568, A77}

\bibitem[\protect\citeauthoryear{{Zapatero Osorio} et~al.,}{{Zapatero Osorio}
  et~al.}{2014c}]{zapatero14c}
{Zapatero Osorio} M.~R.,  et~al., 2014c, \mn@doi [A\&A]
  {10.1051/0004-6361/201424634}, \href
  {http://cdsads.u-strasbg.fr/abs/2014A%26A...572A..67Z} {572, A67}

\bibitem[\protect\citeauthoryear{{de Bruijne}, {Hoogerwerf}, {Brown}, {Aguilar}
   \& {de Zeeuw}}{{de Bruijne} et~al.}{1997}]{deBruijne97}
{de Bruijne} J.~H.~J.,  {Hoogerwerf} R.,  {Brown} A.~G.~A.,  {Aguilar} L.~A.,
  {de Zeeuw} P.~T.,  1997, in ESA SP-402: Hipparcos - Venice '97. pp 575--578

\bibitem[\protect\citeauthoryear{{de Zeeuw}, {Hoogerwerf}, {de Bruijne},
  {Brown}  \& {Blaauw}}{{de Zeeuw} et~al.}{1999}]{deZeeuw99}
{de Zeeuw} P.~T.,  {Hoogerwerf} R.,  {de Bruijne} J.~H.~J.,  {Brown} A.~G.~A.,
   {Blaauw} A.,  1999, AJ, \href
  {http://cdsads.u-strasbg.fr/cgi-bin/nph-bib_query?bibcode=1999AJ....117..354D&amp;db_key=AST}
  {117, 354}

\makeatother
\end{thebibliography}



\appendix

\section{Tables with list of candidates}
\label{USco_VISTA_deepY:Appendix_Tables}

%
%
\begin{table*}
 \centering
 \caption[]{List of USco member candidates that passed our photometric selection criteria ordered by $Y$ magnitude. We list the coordinates in the second epoch deep $Y$ survey (in J2000), optical (deep VISTA $Z$ survey), near-infrared ($Y,J$ from VISTA; $H,K$ from UKIDSS GCS DR9), and mid-infrared ($w1,w2$ from AllWISE) photometry with their error bars and the proper motion in mas/yr of the best photometric candidates. The last part of the table denotes the candidates fainter than the faintest spectroscopic members known in USco newly identified in this work.
 }
{\small
 \begin{tabular}{@{\hspace{0mm}}@{\hspace{0mm}}c @{\hspace{1mm}}c@{\hspace{1mm}}c@{\hspace{1mm}}c@{\hspace{1mm}}c@{\hspace{1mm}}c@{\hspace{1mm}}c@{\hspace{1mm}}c@{\hspace{1mm}}c@{\hspace{1mm}}c@{\hspace{1mm}}c@{\hspace{1mm}}c@{\hspace{0mm}}}
 \hline
 \hline
ID & R.A.\    &     Dec       &  $Y$ & $J$ & $Z$ & $H$ & $K$ & $w1$ & $w2$ & $\mu_{\alpha\cos(\delta)}$ & $\mu\delta$ \cr
 \hline
   & hh:mm:ss.ss & ${^\circ}$:$'$:$''$ & mag &  mag &  mag &  mag &  mag &  mag &  mag &  mas/yr & mas/yr \cr
 \hline
53   & 16:03:50.82 & $-$21:44:56.7 & 15.015$\pm$0.010 & 14.383$\pm$0.018 & 15.757$\pm$0.010 & 13.820$\pm$0.004 &        ---       & 13.226$\pm$0.026 & 13.029$\pm$0.030 &    $-$44.7 &      4.4 \cr
48   & 16:11:17.11 & $-$22:17:17.7 & 15.103$\pm$0.008 & 14.430$\pm$0.009 & 16.118$\pm$0.016 & 13.768$\pm$0.003 & 13.299$\pm$0.003 & 13.065$\pm$0.024 & 12.742$\pm$0.028 &    $-$10.0 &    $-$19.3 \cr
43   & 16:10:30.14 & $-$23:15:17.1 & 15.163$\pm$0.009 & 14.382$\pm$0.007 & 16.194$\pm$0.007 & 13.800$\pm$0.003 & 13.266$\pm$0.003 & 13.069$\pm$0.025 & 12.755$\pm$0.026 &    $-$17.9 &    $-$32.3 \cr
34   & 16:10:00.16 & $-$23:12:19.4 & 15.218$\pm$0.009 & 14.578$\pm$0.010 & 16.093$\pm$0.012 & 13.971$\pm$0.004 & 13.560$\pm$0.004 & 13.391$\pm$0.026 & 13.105$\pm$0.031 &    $-$38.3 &    $-$13.4 \cr
41   & 16:13:30.33 & $-$22:44:06.8 & 15.255$\pm$0.006 & 14.548$\pm$0.010 & 15.937$\pm$0.010 & 13.965$\pm$0.004 & 13.599$\pm$0.004 & 13.467$\pm$0.027 & 13.265$\pm$0.032 &    $-$16.6 &    $-$13.7 \cr
35   & 16:04:40.25 & $-$22:54:32.6 & 15.269$\pm$0.007 & 14.481$\pm$0.019 & 16.486$\pm$0.013 & 13.760$\pm$0.003 & 13.268$\pm$0.003 & 12.816$\pm$0.025 & 12.196$\pm$0.025 &     $-$7.3 &    $-$21.6 \cr
46   & 16:13:40.79 & $-$22:19:46.3 & 15.570$\pm$0.006 & 14.707$\pm$0.007 & 16.580$\pm$0.006 & 14.140$\pm$0.004 & 13.653$\pm$0.004 & 13.416$\pm$0.025 & 13.058$\pm$0.030 &      2.6 &    $-$25.7 \cr
44   & 16:10:51.78 & $-$23:33:27.4 & 15.648$\pm$0.007 & 14.947$\pm$0.006 & 16.460$\pm$0.011 & 14.423$\pm$0.005 & 14.027$\pm$0.005 & 13.819$\pm$0.027 & 13.553$\pm$0.039 &    $-$56.9 &    $-$97.7 \cr
50   & 16:12:53.13 & $-$21:46:27.4 & 15.688$\pm$0.006 & 14.977$\pm$0.020 & 16.335$\pm$0.010 & 14.553$\pm$0.006 & 14.174$\pm$0.006 & 13.975$\pm$0.027 & 13.788$\pm$0.039 &    $-$27.0 &      1.4 \cr
32   & 16:08:30.49 & $-$23:35:11.3 & 15.753$\pm$0.010 & 14.850$\pm$0.006 & 16.929$\pm$0.011 & 14.311$\pm$0.005 & 13.746$\pm$0.005 & 13.360$\pm$0.026 & 12.841$\pm$0.029 &     $-$5.4 &    $-$17.8 \cr
51   & 16:07:23.81 & $-$22:11:02.2 & 16.041$\pm$0.007 & 15.133$\pm$0.009 & 17.269$\pm$0.013 & 14.565$\pm$0.006 & 13.989$\pm$0.006 & 13.707$\pm$0.027 & 13.393$\pm$0.033 &      3.5 &    $-$29.5 \cr
38   & 16:05:01.90 & $-$23:21:30.5 & 16.068$\pm$0.012 & 15.148$\pm$0.007 & 17.226$\pm$0.010 & 14.506$\pm$0.004 & 13.968$\pm$0.005 & 13.778$\pm$0.028 & 13.373$\pm$0.035 &    $-$19.1 &    $-$18.4 \cr
33   & 16:09:05.67 & $-$22:45:16.9 & 16.090$\pm$0.006 & 15.204$\pm$0.009 & 17.321$\pm$0.014 & 14.572$\pm$0.006 & 14.014$\pm$0.005 & 13.817$\pm$0.026 & 13.459$\pm$0.033 &     $-$3.0 &     $-$9.1 \cr
55   & 16:04:42.11 & $-$22:34:11.1 & 16.228$\pm$0.010 & 15.416$\pm$0.012 & 17.148$\pm$0.007 & 14.812$\pm$0.007 & 14.362$\pm$0.007 & 14.172$\pm$0.029 & 14.013$\pm$0.046 &    $-$13.1 &    $-$65.5 \cr
31   & 16:08:28.46 & $-$23:15:10.6 & 16.443$\pm$0.007 & 15.490$\pm$0.009 & 17.676$\pm$0.008 & 14.795$\pm$0.007 & 14.138$\pm$0.006 & 13.765$\pm$0.027 & 13.179$\pm$0.030 &    $-$15.8 &    $-$28.0 \cr
54   & 16:06:02.58 & $-$22:02:49.9 & 16.594$\pm$0.007 & 15.554$\pm$0.011 & 17.916$\pm$0.020 & 14.953$\pm$0.007 & 14.340$\pm$0.007 & 14.097$\pm$0.028 & 13.700$\pm$0.041 &    $-$33.0 &     $-$7.8 \cr
56   & 16:06:03.75 & $-$22:19:30.2 & 16.914$\pm$0.013 & 15.865$\pm$0.015 & 18.302$\pm$0.018 & 15.112$\pm$0.008 & 14.440$\pm$0.007 & 13.932$\pm$0.030 & 13.307$\pm$0.033 &     $-$3.7 &    $-$16.4 \cr
52   & 16:08:18.42 & $-$22:32:25.2 & 17.143$\pm$0.011 & 16.140$\pm$0.012 & 18.541$\pm$0.012 & 15.387$\pm$0.011 & 14.676$\pm$0.010 & 14.331$\pm$0.031 & 13.982$\pm$0.046 &      4.1 &     $-$7.3 \cr
49   & 16:13:01.39 & $-$21:42:54.8 & 17.305$\pm$0.009 & 16.390$\pm$0.018 & 18.427$\pm$0.010 & 15.631$\pm$0.014 & 15.070$\pm$0.013 & 14.859$\pm$0.035 & 14.738$\pm$0.069 &    $-$18.1 &    $-$12.7 \cr
37   & 16:04:16.18 & $-$23:26:29.9 & 17.323$\pm$0.013 & 16.434$\pm$0.007 & 18.685$\pm$0.006 & 15.657$\pm$0.012 & 15.043$\pm$0.013 & 14.669$\pm$0.033 & 14.319$\pm$0.056 &    $-$49.7 &    $-$18.9 \cr
47   & 16:10:38.36 & $-$21:51:47.0 & 17.464$\pm$0.005 & 16.316$\pm$0.020 & 18.899$\pm$0.017 & 15.473$\pm$0.012 & 14.781$\pm$0.009 & 14.350$\pm$0.031 & 13.807$\pm$0.040 &     $-$5.6 &    $-$30.6 \cr
45   & 16:16:10.64 & $-$21:51:24.5 & 17.646$\pm$0.005 & 16.555$\pm$0.009 & 18.831$\pm$0.008 & 15.995$\pm$0.020 & 15.451$\pm$0.018 &        ---       &        ---       &    $-$12.6 &    $-$26.7 \cr
29   & 16:07:14.78 & $-$23:21:01.4 & 17.671$\pm$0.008 & 16.581$\pm$0.004 & 19.124$\pm$0.008 & 15.807$\pm$0.016 & 15.059$\pm$0.014 & 14.464$\pm$0.033 & 13.812$\pm$0.041 &      5.5 &    $-$24.1 \cr
30   & 16:07:37.97 & $-$22:42:47.2 & 17.822$\pm$0.011 & 16.819$\pm$0.008 & 19.412$\pm$0.024 & 16.091$\pm$0.020 & 15.450$\pm$0.019 & 15.098$\pm$0.039 & 14.743$\pm$0.075 &     $-$1.1 &    $-$12.1 \cr
36   & 16:06:09.04 & $-$23:38:11.6 & 18.035$\pm$0.006 & 16.911$\pm$0.008 & 19.501$\pm$0.012 & 16.247$\pm$0.020 & 15.564$\pm$0.020 & 15.176$\pm$0.044 & 14.895$\pm$0.085 &     $-$7.2 &    $-$13.0 \cr
39   & 16:04:36.56 & $-$23:25:24.1 & 18.256$\pm$0.023 & 17.145$\pm$0.020 & 19.664$\pm$0.018 & 16.348$\pm$0.021 & 15.671$\pm$0.024 &        ---       &        ---       &    $-$12.5 &    $-$27.8 \cr
57   & 16:05:12.06 & $-$21:45:28.4 & 18.358$\pm$0.010 & 17.080$\pm$0.010 & 19.792$\pm$0.012 & 16.403$\pm$0.024 & 15.645$\pm$0.020 &        ---       &        ---       &    $-$11.1 &    $-$23.0 \cr
40   & 16:14:20.44 & $-$22:59:52.5 & 18.463$\pm$0.006 & 17.443$\pm$0.013 & 19.615$\pm$0.016 & 16.753$\pm$0.034 & 16.313$\pm$0.044 & 16.268$\pm$0.086 & 16.585$\pm$0.365 &    $-$44.6 &    $-$59.9 \cr
42   & 16:14:22.56 & $-$23:31:17.9 & 18.658$\pm$0.009 & 17.442$\pm$0.012 & 20.224$\pm$0.023 & 16.830$\pm$0.034 & 16.080$\pm$0.031 &        ---       &        ---       &     10.3 &    $-$20.8 \cr
8    & 16:14:02.87 & $-$23:14:42.7 & 19.094$\pm$0.008 & 17.879$\pm$0.009 & 20.329$\pm$0.012 & 17.215$\pm$0.051 & 16.606$\pm$0.053 & 16.182$\pm$0.074 & 15.860$\pm$0.169 &      6.2 &     $-$7.6 \cr
9    & 16:15:32.35 & $-$22:59:51.0 & 19.109$\pm$0.007 & 17.878$\pm$0.009 & 20.396$\pm$0.020 & 17.240$\pm$0.056 & 16.564$\pm$0.058 & 15.822$\pm$0.055 & 15.639$\pm$0.146 &    $-$20.0 &     29.1 \cr
27   & 16:06:01.43 & $-$22:06:16.3 & 19.193$\pm$0.008 & 18.037$\pm$0.009 & 20.401$\pm$0.032 & 17.319$\pm$0.055 & 16.792$\pm$0.059 & 16.489$\pm$0.092 & 16.162$\pm$0.227 &    $-$23.3 &     10.5 \cr
21   & 16:07:55.40 & $-$22:33:52.5 & 19.223$\pm$0.015 & 18.055$\pm$0.010 & 20.286$\pm$0.028 & 17.522$\pm$0.070 & 16.860$\pm$0.065 &        ---       &        ---       &    $-$33.4 &      4.8 \cr
25   & 16:09:56.36 & $-$22:22:45.9 & 19.278$\pm$0.016 & 17.871$\pm$0.016 & 20.589$\pm$0.016 & 16.915$\pm$0.039 & 16.221$\pm$0.035 &        ---       &        ---       &      4.5 &    $-$37.6 \cr
23   & 16:09:06.61 & $-$21:41:05.3 & 19.293$\pm$0.005 & 18.128$\pm$0.018 & 20.628$\pm$0.021 & 17.398$\pm$0.067 & 16.758$\pm$0.059 & 16.687$\pm$0.106 & 16.123$\pm$0.217 &    $-$25.4 &     $-$3.0 \cr
20   & 16:07:31.61 & $-$21:46:54.5 & 19.523$\pm$0.009 & 18.000$\pm$0.013 & 20.938$\pm$0.034 & 17.004$\pm$0.046 & 16.200$\pm$0.034 & 15.147$\pm$0.040 & 14.596$\pm$0.062 &      6.2 &    $-$12.5 \cr
24   & 16:09:18.67 & $-$22:29:23.9 & 19.534$\pm$0.017 & 17.948$\pm$0.009 & 20.786$\pm$0.013 & 16.974$\pm$0.041 & 16.139$\pm$0.033 & 15.539$\pm$0.051 & 15.317$\pm$0.114 &    $-$33.2 &    $-$16.1 \cr
4    & 16:04:13.03 & $-$22:41:03.4 & 19.553$\pm$0.027 & 18.193$\pm$0.020 & 20.800$\pm$0.050 & 17.235$\pm$0.054 & 16.416$\pm$0.042 & 15.462$\pm$0.049 & 15.254$\pm$0.105 &    $-$17.0 &     $-$7.9 \cr
12   & 16:13:34.65 & $-$21:36:20.3 & 19.644$\pm$0.008 & 18.207$\pm$0.017 & 20.918$\pm$0.017 & 17.111$\pm$0.045 & 16.385$\pm$0.039 & 15.744$\pm$0.053 & 15.369$\pm$0.120 &     34.4 &     11.5 \cr
14   & 16:14:07.56 & $-$22:11:52.3 & 19.705$\pm$0.009 & 18.224$\pm$0.016 & 21.105$\pm$0.020 & 17.386$\pm$0.070 & 16.384$\pm$0.049 & 15.570$\pm$0.051 & 15.290$\pm$0.105 &      3.4 &    $-$23.0 \cr
17   & 16:11:10.15 & $-$21:45:16.9 & 19.745$\pm$0.017 & 18.410$\pm$0.027 & 21.938$\pm$0.144 & 17.303$\pm$0.057 & 16.441$\pm$0.040 &        ---       &        ---       &     $-$0.6 &    $-$30.4 \cr
19   & 16:11:44.37 & $-$22:15:44.7 & 19.822$\pm$0.008 & 17.909$\pm$0.010 & 20.730$\pm$0.027 & 16.871$\pm$0.044 & 15.898$\pm$0.030 & 15.303$\pm$0.045 & 14.726$\pm$0.070 &     $-$1.3 &    $-$16.9 \cr
3    & 16:08:43.43 & $-$22:45:16.2 & 19.850$\pm$0.018 & 18.317$\pm$0.021 & 21.162$\pm$0.041 & 17.192$\pm$0.056 & 16.299$\pm$0.040 & 15.412$\pm$0.045 & 14.670$\pm$0.069 &     $-$4.8 &     $-$6.0 \cr
28   & 16:04:20.42 & $-$21:34:53.1 & 19.899$\pm$0.012 & 18.448$\pm$0.012 & 21.113$\pm$0.022 & 17.432$\pm$0.061 & 16.498$\pm$0.045 & 15.695$\pm$0.051 & 15.332$\pm$0.110 &    $-$27.4 &    $-$14.5 \cr
26   & 16:08:42.37 & $-$22:23:26.3 & 19.909$\pm$0.012 & 18.625$\pm$0.020 & 21.169$\pm$0.025 & 17.870$\pm$0.094 & 17.245$\pm$0.091 & 16.959$\pm$0.149 & 16.118$\pm$0.235 &     $-$8.4 &    $-$60.5 \cr
18   & 16:11:10.83 & $-$22:03:47.6 & 20.081$\pm$0.011 & 18.857$\pm$0.012 & 21.260$\pm$0.027 & 18.047$\pm$0.118 & 17.325$\pm$0.092 &        ---       &        ---       &     16.4 &    $-$23.3 \cr
16   & 16:11:05.28 & $-$22:22:57.4 & 20.466$\pm$0.016 & 19.149$\pm$0.019 & 21.953$\pm$0.046 & 18.461$\pm$0.192 & 17.313$\pm$0.109 &        ---       &        ---       &     $-$8.3 &    $-$11.5 \cr
5    & 16:04:31.51 & $-$23:13:04.4 & 20.549$\pm$0.013 & 19.123$\pm$0.021 & 21.905$\pm$0.046 & 18.464$\pm$0.152 & 17.629$\pm$0.140 &        ---       &        ---       &    $-$31.9 &      6.9 \cr
13   & 16:15:15.22 & $-$21:48:46.6 & 20.922$\pm$0.015 & 19.515$\pm$0.029 & 22.287$\pm$0.051 &        ---       &        ---       &        ---       &        ---       &     21.7 &     $-$7.0 \cr
15   & 16:15:12.70 & $-$22:29:49.2 & 20.943$\pm$0.019 & 19.264$\pm$0.025 & 22.307$\pm$0.071 & 18.167$\pm$0.142 & 17.065$\pm$0.087 & 16.529$\pm$0.100 & 15.729$\pm$0.156 &     $-$7.9 &      1.4 \cr
22   & 16:08:34.28 & $-$21:48:04.9 & 20.984$\pm$0.015 & 19.346$\pm$0.039 & 22.269$\pm$0.094 &        ---       & 17.799$\pm$0.152 & 17.022$\pm$0.153 & 16.939$\pm$0.476 &    $-$10.0 &      3.0 \cr
\hline
11   & 16:12:37.84 & $-$22:51:12.9 & 21.781$\pm$0.038 & 19.504$\pm$0.039 & 22.259$\pm$0.062 &        ---       &        ---       &        ---       &        ---       &     83.9 &   $-$564.2 \cr
6    & 16:05:17.84 & $-$23:35:51.9 & 21.824$\pm$0.019 & 18.064$\pm$0.029 & 21.960$\pm$0.055 &        ---       &        ---       &        ---       &        ---       &   $-$259.8 &    $-$17.4 \cr
1    & 16:09:29.08 & $-$23:40:48.2 & 21.909$\pm$0.136 & 19.336$\pm$0.040 & 22.700$\pm$0.083 & 18.368$\pm$0.168 &        ---       &        ---       &        ---       &   $-$486.5 &    $-$47.5 \cr
7    & 16:04:04.50 & $-$23:21:31.0 & 22.164$\pm$0.049 & 19.688$\pm$0.050 & 22.496$\pm$0.082 &        ---       &        ---       &        ---       &        ---       &   $-$155.2 &   $-$350.4 \cr
10   & 16:15:04.04 & $-$22:40:13.3 & 22.199$\pm$0.097 & 19.556$\pm$0.022 & 22.634$\pm$0.084 & 18.430$\pm$0.170 &        ---       &        ---       &        ---       &    377.2 &   $-$356.6 \cr
2    & 16:10:03.24 & $-$23:18:11.8 & 22.756$\pm$0.078 & 20.444$\pm$0.068 & 23.296$\pm$0.159 &        ---       &        ---       & 17.500$\pm$0.199 & 16.519$\pm$0.000 &   $-$564.3 &    $-$43.1 \cr
\hline
\label{tab_USco_VISTA_deepY:new_candYJ_best}
 \end{tabular}
}
\end{table*}
%

%
%
\begin{table*}
 \centering
 \caption[]{List of USco member $YJ$ candidates undetected in the deep VISTA $Z$-band survey. They are among the 10 newly identified potential candidates fainter than the coolest members knonw in USco. None of these candidates has counterparts in UKIDSS GCS DR9, AllWISE, and PanStarrs. Only the last object in the table has a $Z$ magnitude from the deep Subaru/HSC survey ($Z$\,=\,22.959$\pm$0.025 mag; Table \ref{tab_USco_VISTA_deepY:new_candYJ_noZ}). We give their ID number, coordinates in the second epoch deep $Y$ survey (in J2000), $YJ$ photometry with their error bars, and the proper motion in mas/yr measured between the first ($J$) and second ($Y$) epoch VISTA surveys.
 }
{\small
 \begin{tabular}{@{\hspace{0mm}}@{\hspace{0mm}}c @{\hspace{1mm}}c@{\hspace{1mm}}c@{\hspace{1mm}}c@{\hspace{1mm}}c@{\hspace{1mm}}c@{\hspace{1mm}}c@{\hspace{1mm}}c@{\hspace{1mm}}c@{\hspace{0mm}}}
 \hline
 \hline
ID & R.A.\    &     Dec       &  $Y$ & $J$ & $\mu_{\alpha}\cos\delta$ & $\mu_{\delta}$ \cr
 \hline
   & hh:mm:ss.ss & ${^\circ}$:$'$:$''$  &  mag &  mag &  mas/yr & mas/yr \cr
 \hline
candYJ\_VISTA\_101 & 16:09:13.79 & $-$23:35:46.9 & 22.227$\pm$0.166 & 20.247$\pm$0.089 &   $-$149.5 &    109.0 \cr
candYJ\_VISTA\_102 & 16:04:46.67 & $-$22:52:41.4 & 22.459$\pm$0.085 & 20.367$\pm$0.083 &   $-$272.8 &     65.1 \cr
candYJ\_VISTA\_103 & 16:06:26.99 & $-$22:50:00.6 & 22.183$\pm$0.129 & 20.086$\pm$0.058 &   $-$352.1 &    448.9 \cr
candYJ\_VISTA\_104 & 16:14:41.41 & $-$22:16:29.8 & 22.669$\pm$0.070 & 19.569$\pm$0.046 &   $-$330.8 &    340.0 \cr
\hline
\label{tab_USco_VISTA_deepY:new_candYJ_HSCZ}
 \end{tabular}
}
\end{table*}
%

%
%
\begin{table*}
 \centering
 \caption[]{List of USco member $YJ$ candidates with detections in the Subaru/HSC $Z$-band survey. We give their ID number, coordinates in the second epoch deep $Y$ survey (in J2000), Subaru $Z$ photometry, $YJ$ photometry with their error bars. The two brightest in $Z$ appear as the best candidates because they extend the USco sequence shown in colour-colour diagrams involving the $ZYJ$ filters.
 }
{\small
 \begin{tabular}{@{\hspace{0mm}}@{\hspace{0mm}}c @{\hspace{1mm}}c@{\hspace{1mm}}c@{\hspace{1mm}}c@{\hspace{1mm}}c@{\hspace{1mm}}c@{\hspace{1mm}}c@{\hspace{1mm}}c@{\hspace{0mm}}}
 \hline
 \hline
ID & R.A.\    &     Dec       &  $Z$ & $Y$ & $J$ & \cr
 \hline
   & hh:mm:ss.ss & ${^\circ}$:$'$:$''$  &  mag & mag &  mag  \cr
 \hline
candYJ\_HSC\_1005 & 16:12:03.71 & $-$22:35:09.4 & 23.879$\pm$0.064 & 22.318$\pm$0.077 & 20.536$\pm$0.081 \cr
candYJ\_HSC\_1002 & 16:13:59.13 & $-$22:12:36.8 & 24.192$\pm$0.100 & 23.082$\pm$0.122 & 21.176$\pm$0.114 \cr
candYJ\_HSC\_1001 & 16:14:14.47 & $-$23:12:36.7 & 24.266$\pm$0.096 & 22.961$\pm$0.171 & 21.082$\pm$0.107 \cr
candYJ\_HSC\_1003 & 16:14:41.41 & $-$22:16:29.8 & 22.959$\pm$0.025 & 22.669$\pm$0.070 & 19.569$\pm$0.046 \cr
candYJ\_HSC\_1004 & 16:15:32.45 & $-$22:11:05.4 & 23.994$\pm$0.073 & 22.964$\pm$0.126 & 21.202$\pm$0.126 \cr
\cr
\hline
\label{tab_USco_VISTA_deepY:new_candYJ_noZ}
 \end{tabular}
}
\end{table*}
%


\bsp 
\label{lastpage}
\end{document}